\documentclass[12pt,preprint]{aastex}
\usepackage{amsmath}
\usepackage{graphicx}
\usepackage{color}

\shorttitle{Observing the fine structure of loops with coronal rain with SST}
\shortauthors{P. Antolin \& L. Rouppe van der Voort}

\begin{document}

\title{Observing the fine structure of loops through high resolution spectroscopic observations of coronal rain with the CRISP instrument at the Swedish Solar Telescope}

\author{P. Antolin\altaffilmark{1,3} and L. Rouppe van der Voort\altaffilmark{1}}
\affil{\altaffilmark{1}Institute of Theoretical Astrophysics, University of Oslo, P.O. Box 1029, Blindern, NO-0315 Oslo, Norway}
\altaffiltext{3}{Also at: Center of Mathematics for Applications, University of Oslo, P.O. Box 1053, Blindern, NO-0316, Oslo, Norway}
\email{patrick.antolin@astro.uio.no, v.d.v.l.rouppe@astro.uio.no}

\begin{abstract}

Observed in cool chromospheric lines such as H$\alpha$ or \ion{Ca}{2} H, coronal rain corresponds to cool and dense plasma falling from coronal heights. Considered rather as a peculiar sporadic phenomenon of active regions, it has not received much attention since its discovery more than 40 years ago. Yet, it has been shown recently that a close relationship exists between this phenomenon and the coronal heating mechanism. Indeed, numerical simulations have shown that this phenomenon is most likely due to a loss of thermal equilibrium ensuing from a heating mechanism acting mostly towards the footpoints of loops. We present here one of the first high resolution spectroscopic observations of coronal rain, performed with the CRISP instrument at the Swedish Solar Telescope. This work constitutes the first attempt to assess the importance of coronal rain in the understanding of the coronal magnetic field in active regions. With the present resolution, coronal rain is observed to literally invade the entire field of view. A large statistical set is obtained in which dynamics (total velocities and accelerations), shapes (lengths and widths), trajectories (angles of fall of the blobs) and thermodynamic properties (temperatures) of the condensations are derived. Specifically, we find that coronal rain is composed of small and dense chromospheric cores with average widths and lengths of $\sim370$~km and $\sim1500$~km respectively, average temperatures below 7000~K, displaying a broad distribution of falling speeds with an average of $\sim70$~km~s$^{-1}$ and accelerations largely below the effective gravity along loops. Through estimates of the ion-neutral coupling in the blobs we show that coronal rain acts as a tracer of the coronal magnetic field, thus supporting the multi-strand loop scenario, and acts as a probe of the local thermodynamic conditions in loops. We further elucidate its potential in coronal heating. We find that the cooling in neighboring strands occurs simultaneously in general suggesting a similar thermodynamic evolution among strands, which can be explained by a common footpoint heating process. Constraints for coronal heating models of loops are thus provided. Estimates of the fraction of coronal volume with coronal rain give values between 10~\% and 40~\%. Estimates of the occurrence time of the phenomenon in loops set times between 4 and 15~hours, implying that coronal rain may be a common phenomenon, in agreement with the frequent observations of cool downflows in EUV lines. The coronal mass drain rate in the form of coronal rain is estimated to be on the order of $10^{10}$~g~s$^{-1}$, a significant quantity compared to the estimate of mass flux into the corona from spicules. 

\end{abstract}


\section{Introduction}

Solar observations over the past 30 years have presented a very dynamic picture of the Sun. Loops, closed magnetic field entities encountered basically everywhere in the Sun at different spatial scales (of a few Mm and up to a few hundred Mm) are perfect examples of this dynamic character. Especially in active region coronae, studies of such entities reveal a large range of lifetimes, dynamics, and thermodynamic properties which have led to a main trifold classification as hot, warm and cool loops \citep[see the extensive review by][]{Reale_2010LRSP....7....5R}. Several debates exist over the characteristics of the heating leading to such observed properties. For instance, it is not known whether the heating has mainly an impulsive or a uniform character in time. It is not known whether the heating is generally localized at specific atmospheric heights or is rather uniform over the loop. And, maybe most importantly, the main agent converting the magnetic energy into thermal energy is not known \citep[generally known as AC or DC mechanism depending on its timescale,][]{Klimchuk_2006SoPh..234...41K}. 

Several observational studies have shown, however, that a significant fraction of loops in active regions are out of hydrostatic equilibrium \citep{Aschwanden_2001ApJ...550.1036A, Aschwanden_2001ApJ...559L.171A, Winebarger_etal_2003ApJ...587..439W, Schmieder_etal_2004ApJ...601..530S, Aschwanden_2009ApJ...695...12A}. This family of loops shows in general temperatures below $2-3\times10^{6}~$K and show significant changes over relatively short timescales as compared to hotter loops.  Indeed, heating and cooling processes seem to occur continuously in these loops \citep{Kjeldseth_Brekke_1998SoPh..182...73K, UgarteUrra_etal_2006ApJ...643.1245U, Warren_2007PASJ...59S.675W, Ugarte-Urra_etal_2009ApJ...695..642U, Landi_etal_2009ApJ...695..221L, Dudik_etal_2011AA...531A.115D}. A recurrent finding in these and other active region studies is the presence of cool downflows along the legs of the loops \citep{Foukal_1976ApJ...210..575F, Foukal_1978ApJ...223.1046F, Schrijver_2001SoPh..198..325S, Oshea_etal_2007AA...475L..25O, Tripathi_etal_2009ApJ...694.1256T}.  This non-equilibrium state is generally explained through footpoint heating, a scenario in which the loops have their main heating source located towards the footpoints.

Footpoint heating in active region loops has received further observational support. Upflows often appear correlated with nonthermal velocities at the footpoints of these loops \citep{Doschek_etal_2007ApJ...667L.109D, Hara_2008ApJ...678L..67H, Nishizuka_Hara_2011ApJ...737L..43N}. Recently, \citet{Hansteen_2010ApJ...718.1070H} and \citet{DePontieu_etal_2011Sci...331...55D}  have shown that a considerable part of the hot coronal plasma could be heated at low spicular heights, thus explaining the fading character of the ubiquitous type II spicules \citep{Rutten_2006ASPC..354..276R, DePontieu_etal_2007PASJ...59S.655D, Rouppe_etal_2009ApJ...705..272R}. But perhaps one of the clearest evidences for footpoint heating is put forward by the presence of cool structures in the active region coronae, such as filaments/prominences and coronal rain.

Coronal rain is not a newly observed phenomenon. Although observed for the first time almost 40 years ago \citep{Kawaguchi_1970PASJ...22..405K, Leroy_1972SoPh...25..413L} few observational work exists on the subject. Furthermore, the term `coronal rain' is often erroneously attributed to basically any cold material falling down from coronal heights, as has been the case for describing prominence material falling down following a prominence eruption. Coronal rain corresponds to cool and dense blob-like material forming in the hot coronal environment in a timescale of minutes, which subsequently falls down to the surface along loop-like paths. The temperatures of the plasma composing coronal rain seem to range from transition region to chromospheric, according to the lines in which it has been observed: in \ion{Ne}{7} 465~\AA\ and \ion{O}{6} 1032~\AA\ with {\it Skylab} \citep{Levine_Withbroe_1977SoPh...51...83L}, in \ion{O}{5} 629~\AA\ with {\it SoHO}/CDS \citep{Kjeldseth_Brekke_1998SoPh..182...73K}, in the 1600~\AA\ channel of {\it TRACE} \citep{Schrijver_2001SoPh..198..325S}, in the 304~\AA\ channel of {\it SoHO}/EIT \citep{DeGroof_2004AA...415.1141D, DeGroof05, Stenborg_etal_2008ApJ...674.1201S}, in the 304~\AA\ channel of {\it SDO}/AIA \citep{Kamio_etal_2011AA...532A..96K}, in \ion{Ca}{2}~H with {\it Hinode}/SOT \citep{Antolin_2010ApJ...716..154A, Antolin_Verwichte_2011ApJ...736..121A}, in H$\alpha$ \citep[at Okayama Astrophysical Observatory, the Observatoire du Pic du Midi, Big Bear Observatory, the Swedish Vacuum Solar Telescope (SVST):][respectively]{Kawaguchi_1970PASJ...22..405K, DeGroof_2004AA...415.1141D, Leroy_1972SoPh...25..413L, Muller_2005ESASP.596E..37M}.

\citet{Schrijver_2001SoPh..198..325S} followed the various stages of coronal rain formation in loops with the channels of {\it TRACE}. The hot loops rapidly cool down through thermal conduction and radiation until becoming thermally unstable. These radiative instabilities occur locally in the corona and lead to the rapid formation of blob-like condensations. These first appear in the 1216~\AA\ channel and then in the 1600~\AA\ channel suggesting continuous cooling below transition region temperatures. \citet{DeGroof05} analyzed coronal rain simultaneously with {\it SOHO}/EIT in the 304~\AA\ channel and in H$\alpha$ images from Big Bear Observatory. The blobs first appear in EIT before becoming bright in H$\alpha$, supporting the progressive cooling scenario. Furthermore, \citet{Schrijver_2001SoPh..198..325S} shows that the loops hosting the condensations bright in the 1216~\AA\ were still visible in the 171~\AA\ channel, suggesting a multi-temperature structure in the loops. Recently, \citet{Kamio_etal_2011AA...532A..96K} in observations with {\it SDO}/AIA, similarly report cool condensations bright in the 304~\AA\ channel that appear to be surrounded by diffuse hotter plasma in the 171~\AA\ channel.

Numerical simulations have pinpointed the now generally accepted mechanism behind coronal rain. Termed thermal non-equilibrium (or catastrophic cooling) the mechanism is based on the fact that a coronal loop whose heating input is mostly concentrated towards the footpoints will undergo a thermal instability locally in the corona. Due to the footpoint heating source the loop gets rapidly hot and dense. Thermal conduction acts efficiently for slowing the temperature increase over the loop. The heating source together with chromospheric evaporation will gradually increase the density in the loop bringing the heating per unit mass down and with it the average coronal temperature. Eventually, in a time scale of an hour, the temperature is sufficiently low to allow recombination of atoms locally, decreasing the temperature and the pressure locally in a timescale of minutes. The pressure drop accretes mass from the surroundings and produces upflows from the footpoints leading to the condensations. The dense clumps cool rapidly through radiation until the plasma becomes optically thicker and the cooling slows down \citep{Goldsmith_1971SoPh...19...86G, Hildner_1974SoPh...35..123H, Mok_etal_1990ApJ...359..228M, Antiochos_Klimchuk_1991ApJ...378..372A, Antiochos_1999ApJ...512..985A, Karpen_etal_2001ApJ...553L..85K, Muller_2003AA...411..605M, Muller_2004AA...424..289M, Muller_2005AA...436.1067M, Mendozabriceno_2005ApJ...624.1080M, Mok_etal_2008ApJ...679L.161M, Antolin_2010ApJ...716..154A, Xia_etal_2011ApJ...737...27X}. It is still not clear how cool can coronal rain become from this process (before falling down to the solar surface).

Since they have so far been observed in the same spectrum lines, coronal rain is often associated with prominences, and probably for both the thermal non-equilibrium mechanism is equally important \citep{Dahlburg_etal_1998ApJ...495..485D, Karpen_etal_2001ApJ...553L..85K, Karpen_etal_2005ApJ...635.1319K, Karpen_etal_2006ApJ...637..531K, Liu_etal_11SPD....42.2119L}. However, one fundamental difference seems to be the magnetic field structure in which they appear, most likely determining the time scale of the phenomenon. While a prominence can live for days or even weeks, coronal rain can only exist during the time it takes for the blobs to fall to the surface (on the order of half an hour or less). The thermodynamic properties of the plasma are very likely linked to the underlying magnetic field topology. Coronal rain is restrained to coronal loops, where the magnetic field is basically bipolar and magnetic field lines are not far from being potential. On the other hand, the magnetic field topology in prominences seems to present complicated geometries such as `dips' or `peaks'  \citep[see][and references therein]{Lin_2010SSRv..tmp..112L}. Helical magnetic fields conforming flux ropes have also been proposed, in which apart from offering support from the Lorentz force to balance gravity, thermal isolation from the surrounding hot corona is also achieved due to the cross-field thermal conduction inhibition \citep{Kippenhahn_Schluter_1957ZA.....43...36K, Low_Hundhausen_1995ApJ...443..818L}. The longer timescales involved in prominences may produce different element population leading to different ionization fraction, and consequently mechanisms such as ambipolar diffusion may play a different role. Another type of prominence model has been suggested in which the material is continuously renewed through flows between the prominence and the chromosphere, giving the impression of long-lived material \citep{Priest_Smith_1979SoPh...64..267P, Tandberg-Hansen_1995ASSL..199.....T, Zirker_etal_1998Natur.396..440Z, Liu_etal_11SPD....42.2119L, Karpen_etal_2001ApJ...553L..85K, Karpen_etal_2005ApJ...635.1319K}. Since this case does not involve long-term gravitational support the required magnetic field may be more simple and features such as `dips' are not necessary: conditions that are more alike to coronal rain formation. 

High resolution observations have revealed that prominences are often composed of a myriad of fine threads, outlining a fine-scale structure of the magnetic field and the presence of flows along the threads \citep{Heinzel_Anzer_2006ApJ...643L..65H, Lin_2005SoPh..226..239L, Lin_2008ASPC..383..235L, Lin_2010SSRv..tmp..112L, Martin_2008SoPh..250...31M}. Observations of coronal rain with {\it Hinode}/SOT have also revealed a thread-like character in coronal loops, due to the separation and elongation of the blobs down to very small sizes \citep{Antolin_2010ApJ...716..154A, Antolin_Verwichte_2011ApJ...736..121A}. The observations with the {\it SST} presented here allow a closer look into this thread-like character of loops, suggesting the actual tracing of strands by the blobs. 

Due to the strong footpoint heating and the subsequent very high densities that follow a flare, thermal non-equilibrium leading to the formation of condensations is also proposed as the mechanism behind the observed cool downflows in post-flare loops  \citep{Foukal_1978ApJ...223.1046F, Schmieder_etal_1995SoPh..156..337S, Shimojo_etal_2002SoPh..206..133S, Hara_etal_2006ApJ...648..712H}. Despite the very different thermodynamic conditions in post-flare loops with respect to warm loops, the coronal rain observed in both cases appears similar. Although a rigorous comparison awaits, this could be easily explained since radiative cooling increases with the square of the density, and therefore achieving the same low temperatures in a similar timescale. It is possible however that the element population of the obtained blobs in post-flare loops is different, leading to different properties at short length scales. It is thus important to make the distinction between flaring and non-flaring active regions for coronal rain studies. 

Coronal rain is often observed to follow loop-like paths, suggesting a strong coupling between neutrals coming from recombination and the local ion population. Since the thermodynamical state and ionization fraction of such clumps are still poorly known it is not clear how strong this coupling is and, consequently, how well coronal rain blobs can trace the coronal magnetic field. Assuming that a blob does not alter the magnetic field lines, by studying its dynamics, morphology changes and thermodynamics we can learn about the local conditions in coronal loops. This is an idea that is exploited in this work. Supported by theoretical estimates of coupling processes based on the obtained results we show that the very small sizes of blobs allow them to be magnetic field tracers and probes of the local environment. 

In all of the observational papers concerning coronal rain it has been shown that the falling speeds are smaller than free fall, leading to downward accelerations considerably lower than that of the solar gravity at the surface. The gas pressure gradient inside the loops has often been suggested as the main agent regulating the speeds of the blobs. This is also supported by numerical simulations of thermal non-equilibrium \citep{Mackay_2001SoPh..198..289M, Muller_2003AA...411..605M}. Magnetic pressure exerted by transverse MHD waves have also been suggested as factors determining the kinematics of blobs. In \ion{Ca}{2}~H observations with {\it Hinode}/SOT, \citet{Antolin_Verwichte_2011ApJ...736..121A} detected such waves through coronal rain tracking. It is shown that the pressure force from the waves can explain the low falling speeds. Further investigation on the effect of waves on blobs such as those of coronal rain is being carried out \citep{Verwichte_etal_2011}. Whether blobs are capable of producing such waves is also being investigated in that work.

Based on the {\it TRACE} observations, \cite{Schrijver_2001SoPh..198..325S} estimated the occurrence time of coronal rain in a coronal active region loop to be at most once every two days stressing the sporadic character of coronal rain in active regions, and thus suggesting that thermal non-equilibrium may not play a significant role in coronal heating. These estimates are constrained by the resolution of {\it TRACE}. Here we show that in order to have a clear picture about the phenomenon a resolution at least twice that of {\it TRACE} is needed. New estimates of the occurrence time of coronal rain and the coronal volume involved are presented, as well as the coronal mass drain rate. 

The importance of thermal non-equilibrium (catastrophic cooling) in coronal heating is a matter of significant debate. So far, observations have pointed out five characteristics of warm loops ($\sim1~$MK) in active regions, which any coronal heating model must explain. Namely, an excess density, a flat temperature profile, a super-hydrostatic scale height, an unstructured intensity profile, and a $1000-5000~$s lifetime. \cite{Klimchuk_2010ApJ...714.1239K} have analyzed through one-dimensional simulations the case of thermal non-equilibrium in this context, considering both monolithic and multi-stranded loops. The simulations failed in general to satisfactorily reproduce the observations. For instance, monolithic models end up with far too much intensity structure, while multi-strand models are either too structured or too long-lived. On the other hand, 3D-MHD models of an active region with a footpoint heating function leading to thermal non-equilibrium have provided satisfactory results \citep{Mok_etal_2008ApJ...679L.161M}. The resulting evolution of the light-curves, the variation of temperature along the loops, the density profile, and the absence of small scale structures in the intensity profiles are all compatible with real loops, thus supporting the thermal non-equilibrium scenario \citep{Lionello_etal_2010AGUFMSH31C1811L}. 

Impulsive versus steady heating of loops is a common debate in studies of coronal heating \citep[see][and references therein]{Klimchuk_2006SoPh..234...41K}. The scenario in which loops are heated impulsively to hot temperatures and subsequently cool down to EUV or lower temperatures is often proposed in observational studies of the corona \citep{Warren_etal_2002ApJ...579L..41W, Winebarger_etal_2003ApJ...593.1164W, Winebarger_etal_2003ApJ...593.1174W, Viall_Klimchuk_2011ApJ...738...24V} but neither this nor the steady heating scenario are fully compatible with some observations \citep{UgarteUrra_etal_2006ApJ...643.1245U}. Indeed, in some cooling events the {\it TRACE} or {\it SDO}/AIA filters do not seem to brighten according to their temperature. A possible solution to this has been put forward by \citet{Kamio_etal_2011AA...532A..96K}, where, apart from investigating the upward propagation of quasi-periodic brightness fluctuations observed with {\it SDO}/AIA, they also analyze sporadic cool downflows that they interpret as coronal rain. The peculiar order of the brightening in the AIA channels in time seem to be due to these cool downflows, particularly, because of the contribution of cool lines to hot channels in the AIA temperature response functions. Throughout their lifetimes the condensations progressively cool and the densities increase (in order to keep the local pressure roughly constant), producing a considerable increase in the cool line emission. The authors thus favor the thermal non-equilibrium mechanism to explain the observational results.

Being the observational signature of thermal non-equilibrium (and hence to footpoint heating), coronal rain is deeply linked with coronal heating. As shown by the numerical simulations of the works cited previously thermal non-equilibrium is very sensitive to parameters such as the heating scale height and the loop length. By studying coronal rain we can then learn about the heating mechanisms. \citet{Antolin_2010ApJ...716..154A} showed that Alfv\'en wave heating, a strong coronal heating candidate, is not a predominant heating mechanism in loops with coronal rain. When propagating from the photosphere into the corona, Alfv\'en waves can nonlinearly convert to longitudinal modes due to density fluctuations, wave-to-wave interaction, and deformation of the wave shape during propagation \citep{Vasheghani_2011AA...526A..80V}. These modes subsequently steepen into shocks and heat the plasma uniformly along the loop \citep{Moriyasu_2004ApJ...601L.107M, Antolin_2010ApJ...712..494A}, thus avoiding the loss of thermal equilibrium in the corona. In this work we deepen the link between coronal rain and coronal heating by studying the occurrence character of thermal non-equilibrium in neighboring strands. 

This paper is organized as follows. In section~\ref{observations} we present the {\it SST}/CRISP observations. In section~\ref{results} the results of the statistical analysis are presented. These comprise dynamics, shapes, trajectories and thermodynamic properties. Discussion is presented in section~\ref{discussion}, where we try to assess the role of coronal rain in the understanding of the coronal magnetic field. Conclusions are presented in section~\ref{conclusions}.

\section{Spectroscopic observations of coronal rain with CRISP of \textit{SST}}\label{observations}

\subsection{Observations and data reduction}\label{reduc}

The present observations were carried out with the CRisp Imaging SpectroPolarimeter \citep[CRISP;][]{Scharmer_2008ApJ...689L..69S} at the Swedish 1-m Solar Telescope \citep[SST;][]{Scharmer_2003SPIE.4853..341S} sampling the H$\alpha$ spectral line at 25 line positions with 0.01~nm steps, in the spectral window [$-0.12, 0.12$] nm ($\simeq\pm55$~km~s$^{-1}$). The target is shown in Fig.~\ref{fig1} in H$\alpha$ line center and in Fig.~\ref{fig2} in the extreme blue wing of $-0.12~$nm. It covered a plage region containing a few pores on the east solar limb on 10 May 2009, a few days later classified as AR 11017. The observation sequence runs from 08:50 to 10:15 UT with a cadence of 6.36~s, and can be divided into 2 data sets with slightly different pointing: the first, from 08:50 to 09:37 UT centered roughly at (x,y)$\approx$(-900,300), and the second slightly further to the limb, from 09:38 to 10:15 UT. Figure~\ref{fig1} shows a composite image from the two data sets, resulting in an entire field of view of 48$\times 45$~Mm.

CRISP is a spectropolarimeter that includes a dual Fabry-P\'erot interferometer (FPI) system similar to that described by \citet{Scharmer_2006AA...447.1111S}. It is equipped with three high-speed low-noise CCD cameras that operate at a frame rate of 35 frames per second and an exposure time of 17 ms. The three cameras are synchronized by means of an optical chopper, two of these cameras are positioned behind the FPI after a polarizing beam splitter, and the third camera is positioned before the FPI but after the CRISP prefilter. The latter camera is used as anchor channel for image processing and is referred to as the wide-band channel. The image scale pixel is 0$.\!\!^{\prime\prime}$0592 pixel$^{-1}$, and the field of view is about $59\times59$ arcsec. CRISP allows for fast wavelength tuning ($\lesssim50~$ms) within a spectral region and is ideally suited for spectroscopic imaging of the chromosphere where the dynamical evolution time can be on the order of a few seconds, sometimes even faster than 1 s \citep{VanNoort_Rouppe_2006ApJ...648L..67V}. For H$\alpha$ the transmission FWHM of CRISP is 6.6~pm, and the prefilter is 0.49~nm.

The image quality of the time sequences benefited from the \textit{SST} adaptive optics system \citep{Scharmer_etal_2003SPIE.4853..370S} and the image restoration technique Multi-Object Multi-Frame Blind Deconvolution \citep[MOMFBD;][]{VanNoort_etal_2005SoPh..228..191V}. Although the observations suffered from seeing effects, most of the images are close to the theoretical diffraction limit for the \textit{SST} at the wavelength of H$\alpha$:  $\lambda/D\simeq0.\!\!^{\prime\prime}$14. Eight exposures per wavelength position were recorded and individually divided in overlapping 64$\times$64 pixel subfields for processing as a single MOMFBD restoration. In such a restoration, the wide-band channel served as anchor for the narrowband CRISP exposures to ensure precise alignment between the restored narrowband images. We refer to \citet{VanNoort_Rouppe_2008AA...489..429V} for more details on the MOMFBD processing strategies on similar extensive multi-wavelength scans.

After MOMFBD reconstructions, each spectral scan was corrected for the effect of the transmission profile of the CRISP pre-filter (FWHM $0.49~$nm). The correction was determined from a comparison with a solar atlas H$\alpha$ spectral profile. The images from the different time steps were then combined to form time series. The images were de-rotated to account for diurnal field rotation, and rigidly aligned using the reconstructed wide-band images to determine the spatial offsets between time steps which were subsequently applied to the associated CRISP images. 

After the MOMFBD process small misalignments between line positions can still remain. The correction for this spatial mismatch is normally performed through the de-stretching procedure, which consists in first dividing the image into small subfields and using background photospheric structure to align all line positions. The background structure is however absent for off-limb subfields, and thus this procedure could not be applied for the present data. The small misalignments are considered, however, not important for the present work.

The present  data constitute one of the first CRISP off-limb observations where successful AO locks (adaptive optics) were achieved, allowing the high spatial resolution in the images. In order to clearly discern the off-limb features relative to the much brighter disc structure, a radial filter was applied. Faint structures such as coronal rain can then be easily detected.

\subsection{Coronal rain}\label{rain}

\begin{figure}
\epsscale{1.}
\plotone{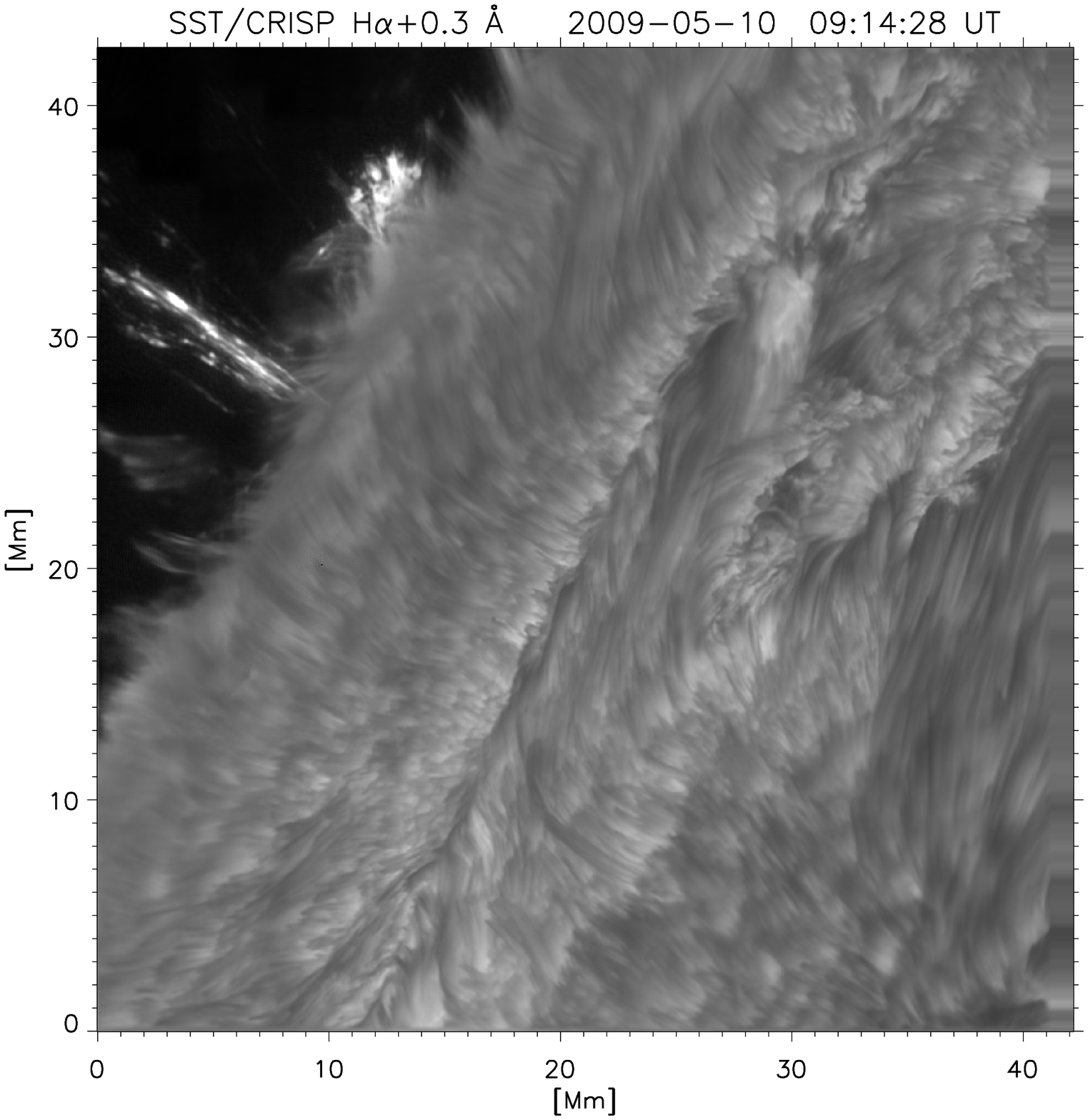}
\caption{Limb observations in H$\alpha$ with CRISP at the {\it SST} of active region 11017 on May 10, 2009. The image corresponds to an offset of 0.3~\AA\ into the red wing of H$\alpha$. In order to enhance the off-limb faint features with respect to the bright disc a radial filter has been applied. The bright clumped material falling down diagonally from the top left corresponds to a group of condensations. This coronal rain event is an example of a `shower'. 
\label{fig1}}
\end{figure}

\begin{figure}
\epsscale{1.}
\plotone{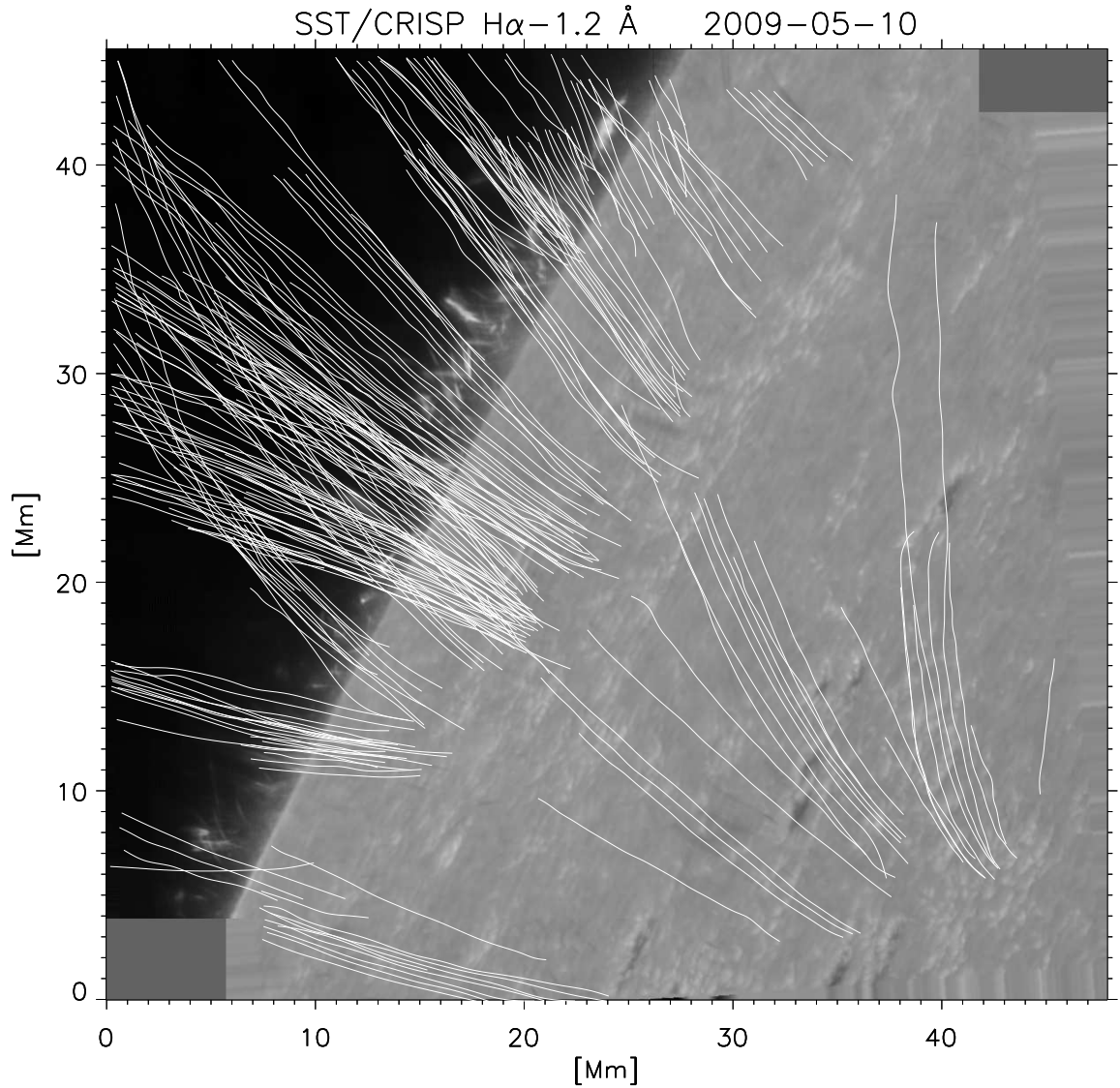}
\caption{Composite image covered by both datasets displaying the full field of view. The offset is $-1.2~$\AA\ towards the blue wing of H$\alpha$. The white curves denote the paths of the tracked condensations. A total of 242 paths with more than 2500 condensations in total were tracked. Note that on-disc events were also spotted (these present absorption profiles).
\label{fig2}}
\end{figure}

During the entire observational sequence bright blobs in different spectral offsets from H$\alpha$ line center are observed to fall from coronal heights down to the lower atmosphere. The second data set presents more off-limb coronal rain than the first, probably due to the wider off-limb section in the field of view. On the other hand, the first data set presents slightly more on-disc coronal rain than the second. The observed blobs could be traced often to the footpoints of the fibrilar structures observed towards H$\alpha$ center that protrude partly above the limb (see Fig.~\ref{fig1}) and seem to have as roots the bright plage region below, which can be observed towards the wing in Fig.~\ref{fig2}. Most of the blobs cross the \textit{SST} field of view following straight paths. No prominence is observed above the limb, indicating that the H$\alpha$ blobs are not material detaching from a prominence. No flares or other explosive events were recorded. This suggests a formation on short timescales along long coronal loops from thermal non-equilibrium, matching the characteristics of coronal rain. Examples of such condensations can be seen in Fig.~\ref{fig1}, falling in a straight line roughly perpendicular to the solar surface. 

In the available on-line movie the ubiquitous character of coronal rain in this dataset can be seen. In order to locate the limb and the plage region on-disc the movie starts on the red wing of H$\alpha$ at an offset of $+1.1$~\AA\  from line center, which is slowly decreased, finishing at line center. Both datasets are sequentially shown. The Doppler velocity of the blobs as calculated in section~\ref{method} is plotted in color, where red (positive) corresponds to red shift and blue (negative) corresponds to blue shift. Apart from the ubiquitous character of the blobs, note the sporadic appearance of large clumps, actually composed of numerous condensations. The large group of blobs in Fig.~\ref{fig1} is one example of such event. We denote these events as `showers'.

\begin{figure}
\epsscale{1.}
\plotone{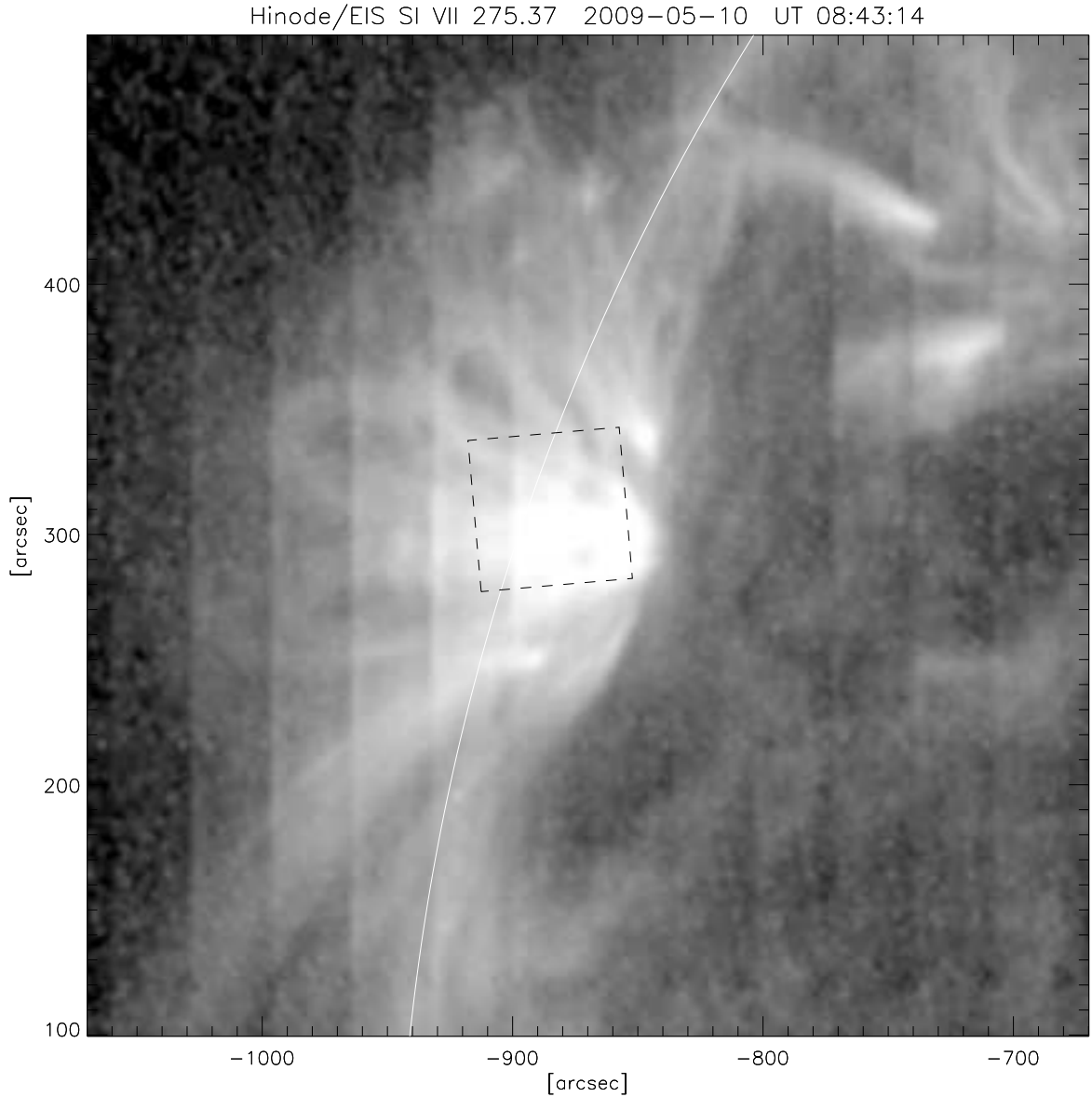}
\caption{{Hinode}/EIS observation in the \ion{Si}{7} 275.37~nm transition region line of the active region 11017 on May 10, 2009. The image corresponds to a $40^{\prime\prime}$ slot reconstruction at 15 adjacent positions, leading to an entire field of view of $487\times487$~arcsecs, of which a subfield is shown. The square in dashed lines corresponds to the approximate position of the {\it SST} field of view shown in Fig.~\ref{fig1}. 
\label{fig3}}
\end{figure}

Figure~\ref{fig3} is an image obtained with the Extreme ultra-violet Imaging Spectrometer \citep[EIS;][]{Culhane_2007SoPh..243...19C} of the \textit{Hinode} satellite \citep{Kosugi_2007SoPh..243....3K} at the wavelength of \ion{Si}{7}~275.37~nm formed at a temperature of $6\times10^{5}$~K. The image corresponds to a $40^{\prime\prime}$ slot reconstruction at 15 adjacent positions, leading to an entire field of view of $487\times487$~arcsecs, of which a subfield is shown. The observations with EIS have a cadence of roughly 3~minutes and run from 08:43 to 09:49 UT, covering most of the {\it SST} observations. The data were reduced using standard EIS software included in the SolarSoft package \citep{Freeland_Handy_1998SoPh..182..497F}. In black dashed lines in the figure we indicate the approximate position of the \textit{SST} field of view. Many loop structures can be observed crossing the \textit{SST} field of view in almost straight curves, thus confirming the picture inferred from the H$\alpha$ observations. These loops seem to be rapidly evolving, several disappearing and others appearing during the course of the observations. However, none of the coronal rain events (neither showers) can be detected (in absorption) in EIS. Most of the loops in the {\it SST} field of view appear to have their baselines directed roughly along the line of sight. A \textit{SOHO}/MDI magnetogram taken a few hours after our observation time shows a bipolar region whose main axis is roughly directed along the line of sight, thus supporting the visual impression from the loop geometries. The loop planes also do not show big inclination angles with respect to the normal to the surface. The apexes of these loops seem to be around $50-70$~Mm above the solar surface, leading to full lengths between 150 and 230~Mm, assuming circular paths. Additionally, longer loops can also be observed interconnecting different plage regions.

The blob-like condensations characterizing coronal rain are observed to change shape easily, increasing or decreasing in size, elongating, separating during their fall, making it sometimes difficult to trace individual blobs all the way down to the footpoints of the loops. For this purpose we used CRISPEX (CRisp SPectral EXplorer) and TANAT (Timeslice ANAlysis Tool) \footnote{The actual code and further information can be found at http://bit.ly/crispex}, two widget based tools programmed in the Interactive Data Language (IDL), which enable the easy browsing of the image and spectral data, the determination of loop paths, extraction and further analysis of space-time diagrams. 

As condensations fall, mostly straight, weakly curving strand-like paths are traced. A total of 242 paths were traced in the 84 minute observational sequence, shown all together in Figure \ref{fig2}. Although projection effects come into play, these paths appear to have the underlying faculae as footpoints. The paths of the blobs seem to match the coronal magnetic field topology shown in Fig.~\ref{fig3}, suggesting that the H$\alpha$ condensations generally trace the inner structure of the coronal loops observed in Fig.~\ref{fig3}, despite their neutral character. We discuss this scenario in section~\ref{discuss1}.

\subsection{Methods}\label{method}

Due to the constant change in shape of the blobs during their fall their tracking was performed manually in order to avoid possible errors that an automated procedure may deliver. Only the clearly discernible blobs were selected. By tracing the blobs during their fall we can construct space-time diagrams such as shown in Figs.~\ref{fig7} and~\ref{fig11}, where the spatial component traces the length along the blob's path. The full spectral profile is then retrieved at each position along the trajectory, from which the Doppler velocity can be calculated. Fitting piece-wise segments along the trajectory we estimate the projected velocity at different locations along the path, and the resulting acceleration. With this method, the standard deviation for projected velocities and accelerations is estimated to be $ 5.1~$km~s$^{-1}$ and $0.03~$km~s$^{-2}$ respectively. 

The calculation of the Doppler velocity of the condensations requires subtraction of the average spectral profile of the background, especially for the cases having a bright chromospheric or photospheric background. Given the path of a condensation, its spectral profile at a specific time was calculated with the help of time slice diagrams. By tracing the $x-t$ trajectory of condensations in diagrams such as those in Fig.~\ref{fig7}, the spectral profile at a specific time and position along the trajectory is calculated averaging over an interval along the trajectory containing the point of interest. The average spectral profile of the background was calculated in the same way but over a time interval in which the trajectory does not exhibit any coronal rain. Only measurements (specific segments in the $x-t$ diagram) with a high enough intensity contrast between the blob and the average background were then selected. The Doppler velocity for a blob at a given position and time was calculated with the first moment with respect to wavelength, a method used by \citet{Rouppe_etal_2009ApJ...705..272R} to calculate the velocities of the disk-counterparts of type II spicules:

\begin{equation}
v_{{\rm Doppler, 1}} = \frac{c}{\lambda_{0}}\frac{\int_{\lambda_{{\rm min}}}^{\lambda_{{\rm max}}}(\lambda-\lambda_{0})|I_{\lambda}-I_{\lambda, {\rm avg}}| d\lambda}{\int_{\lambda_{{\rm min}}}^{\lambda_{{\rm max}}}|I_{\lambda}-I_{\lambda, {\rm avg}}|d\lambda},
\end{equation}
where $c$ is the velocity of light, $\lambda_{0}$ the wavelength at line center, $I_{\lambda}$ the calculated spectral profile of the condensation and $I_{\lambda, {\rm avg}}$ the average spectral profile of the background. The integration range is set by the minimum and maximum wavelengths for which $I_{\lambda}-I_{\lambda, {\rm avg}}>0$ for emission profiles, and $I_{\lambda}-I_{\lambda, {\rm avg}}<0$ for absorption profiles. In order to estimate the error involved in this calculation we performed the integration varying both end points by various amounts $dl$: $\lambda_{{\rm min}}+dl$, $\lambda_{{\rm max}}-dl$. The Doppler velocity with this method was then set as the mean over the resulting values, and the standard deviation gives us an estimation of the error. 

The obtained velocities were checked with two other methods. The first involves a single gaussian fit of $I_{\lambda}-I_{\lambda, {\rm avg}}$ for an emission profile (and $I_{\lambda, {\rm avg}}-I_{\lambda}$ for an absorption profile). The wavelength interval where the gaussian fit is made is basically the same as the $[\lambda_{{\rm min}}, \lambda_{{\rm max}}]$ range defined above and is allowed to vary in the same way in order to estimate the errors involved. The Doppler velocity with this method is taken as $v_{{\rm Doppler, 2}}=\frac{c}{\lambda_{0}}(\lambda_{{\rm gauss}}-\lambda_{0})$, where $\lambda_{{\rm gauss}}$ is the maximum of the gaussian fit. The second check is simply $v_{{\rm Doppler, 3}}=\frac{c}{\lambda_{0}}(\lambda_{{\rm max}}-\lambda_{0})$, where $\lambda_{{\rm max}}$ is the location of the maximum of $|I_{\lambda}-I_{\lambda, {\rm avg}}|$ in the range $[\lambda_{{\rm min}}, \lambda_{{\rm max}}]$. While all three methods deliver similar results we note that the Gaussian method generally delivers slightly lower velocity magnitudes, and the maximum method normally delivers larger values. 

In panel $b$ of Fig.~\ref{fig4} the histogram in black denotes the number of cases having a standard deviation larger than 5~km~s$^{-1}$, where the standard deviation is the maximum of the standard deviations obtained with methods 1 and 2. We can clearly see that the errors involved in the Doppler velocity calculation are generally small. 

For the determination of the heights of the blobs above the solar surface the projected distance of each pixel in the image from disc center was first measured. By tracing a blob down to the lower atmosphere the location of the footpoint of the strand was estimated, and its distance to disc center was determined. The difference in the two measured distances was taken as the height for the blob. Since this method is strongly sensitive to projection effects the heights were only estimated for the blobs observed off-limb. Another source of error is the correct determination of the falling location of the blobs. Figure~\ref{fig2} shows a good correspondence between the later and the bright faculae. We have estimated the error in heights to be $\sim\pm2$~Mm. 

The lengths of the blobs were calculated in the following way. Given a blob at a particular time and spatial position we first select the wavelength position of maximum brightness. We then subtract to the intensity profile along the blob's path the average spectrum of the background corresponding to that path, time and wavelength position, where the average spectrum is calculated as explained previously. Next we fit a gaussian to the resulting intensity profile along the blob's path. We take the FWHM value as a measure of the length at that time. The procedure is repeated along the blob's path providing averages and standard deviations for the lengths of the blob at different heights. The method for calculating widths is similar, with the difference that the gaussian fit is performed over the orthogonal direction to the path of the blob (selecting again the wavelength position of maximum blob brightness at each time).

\section{Results}\label{results}

\subsection{Dynamics}\label{dyna}

\begin{figure}[!ht]
\begin{center}
$\begin{array}{c@{\hspace{-0.2in}}c@{\hspace{-0.2in}}c}
\includegraphics[scale=0.35]{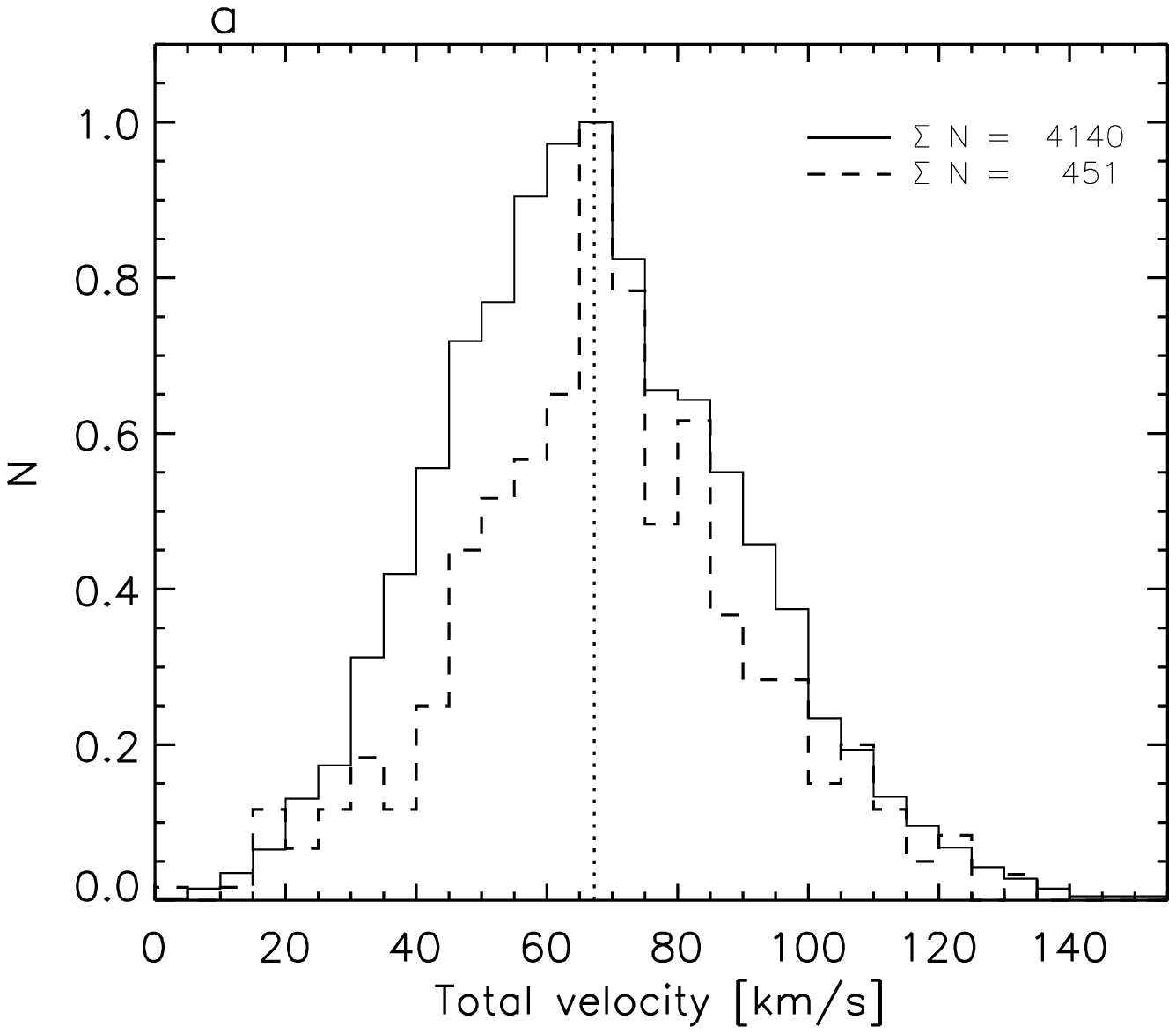}  &  
\includegraphics[scale=0.35]{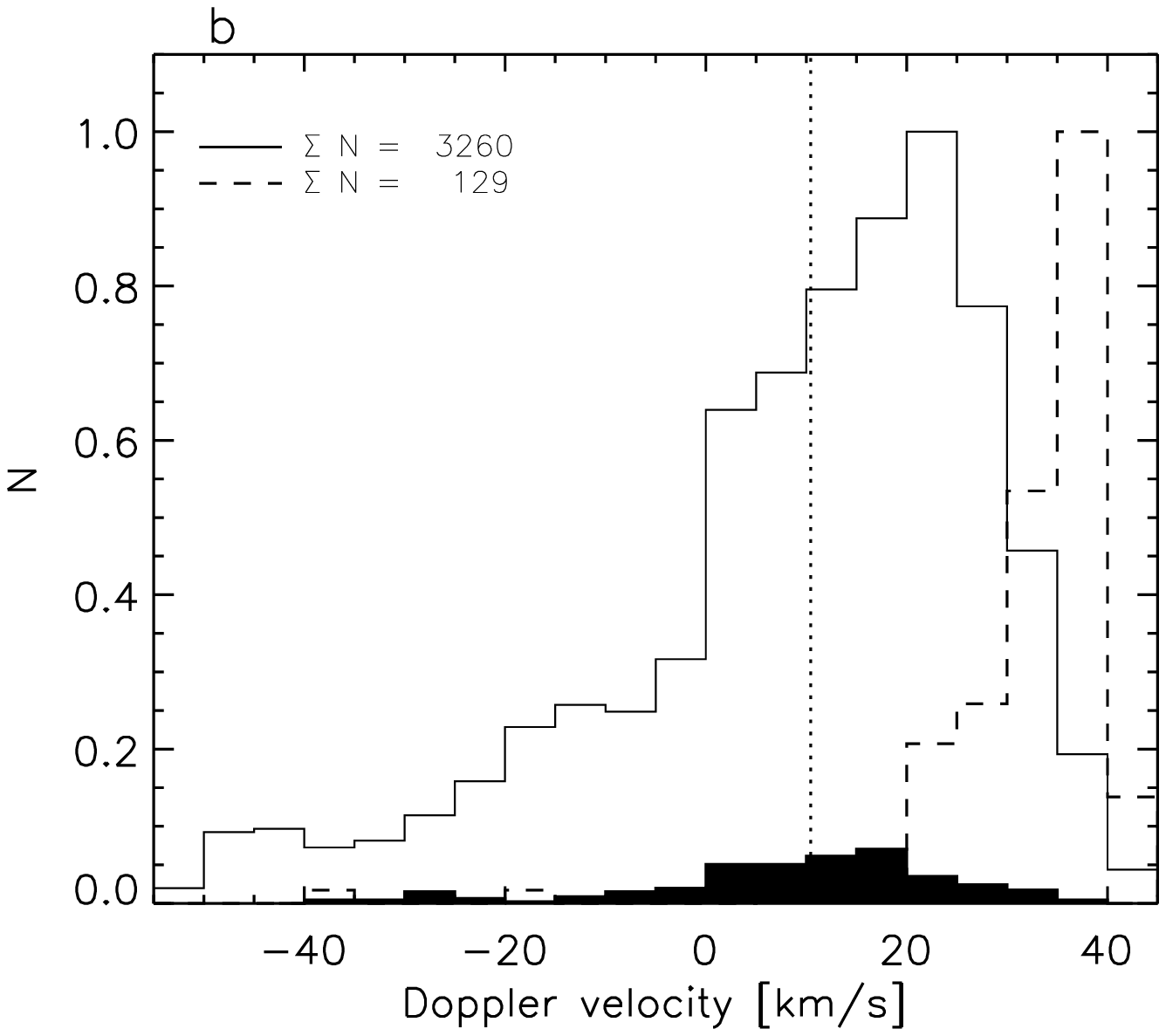} &
\includegraphics[scale=0.35]{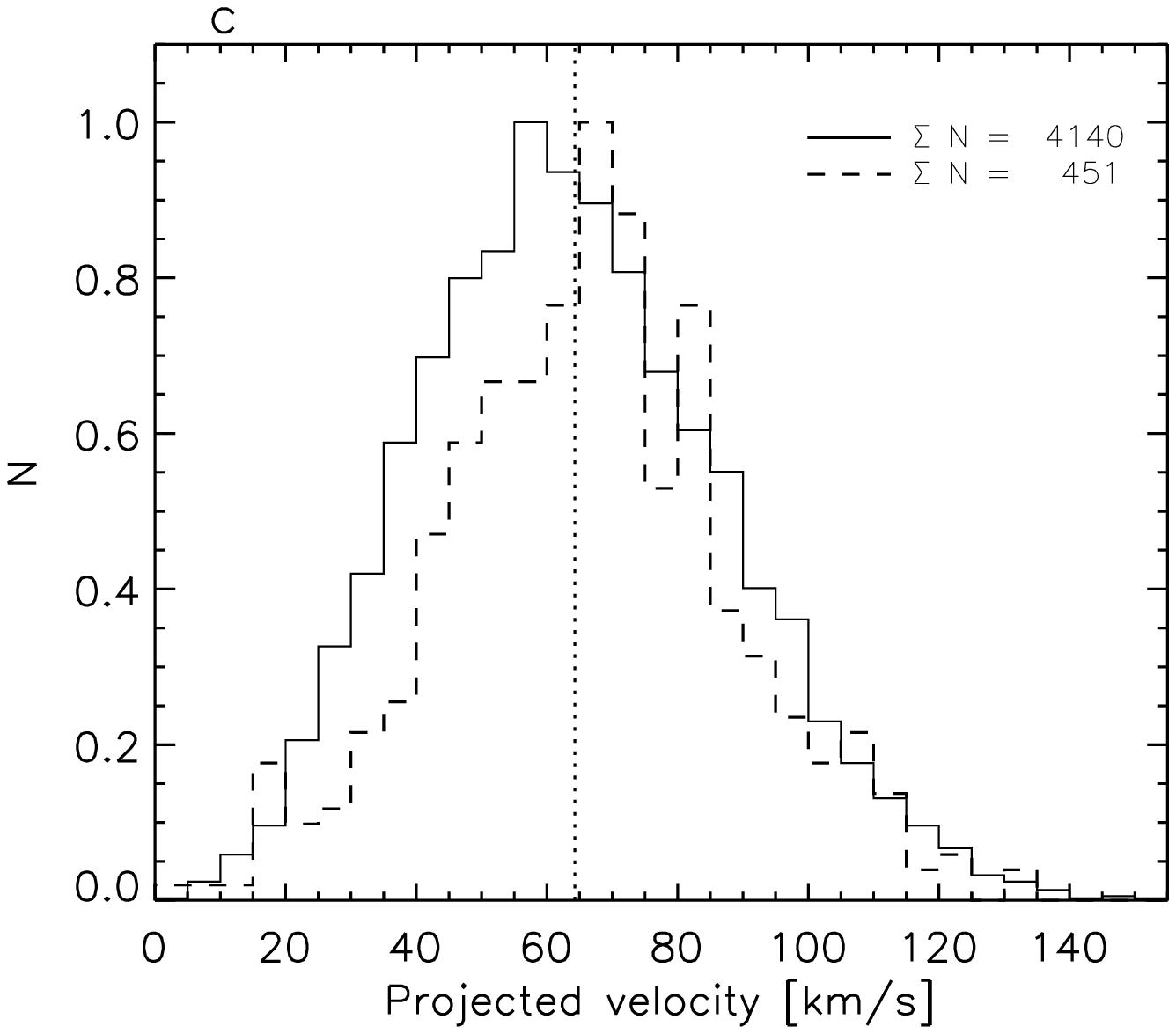}   \\
\end{array}$
\caption{Normalized histograms for the total velocity $v_{{\rm tot}}$ (panel {\it a}), Doppler velocity $v_{{\rm Dop}}$ (panel {\it b}) and projected velocity $v_{{\rm proj}}$ (panel {\it c}) of the condensations, where $v_{{\rm tot}}=\sqrt{v_{{\rm Dop}}^2+v_{{\rm proj}}^2}$. The solid and dashed lines correspond to off-limb and on-disc blobs respectively. The total number of measurements is specified in each panel. The dotted line corresponds to the average over all measurements. The black histogram in panel {\it b} denotes the measurements that have a standard deviation larger than 5~km~s$^{-1}$. See section~\ref{method} for more details. 
\label{fig4}}
\end{center}
\end{figure}

\begin{figure}
\epsscale{1.}
\plotone{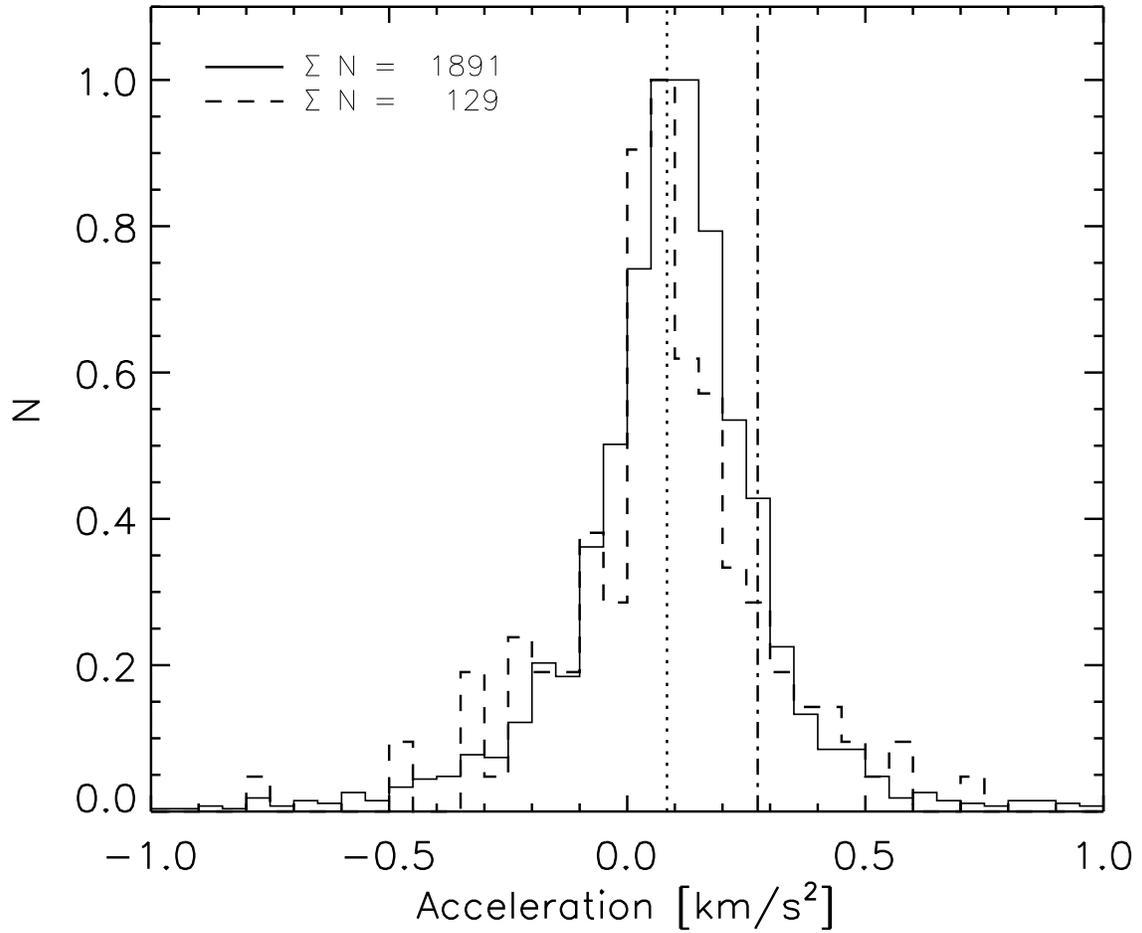}
\caption{Normalized histogram for the acceleration of condensations. The solid and dashed lines correspond to off-limb and on-disc blobs respectively. The total number of measurements is specified in the top left corner. The dotted line corresponds to the average over all measurements and the dot-dashed line corresponds to the solar gravity value at the solar surface, 0.274~km~s$^{-2}$. For further details see section~\ref{method}.
\label{fig5}}
\end{figure}

Many blobs are observed falling along one path. Selecting the most clearly discernible cases, a total of 2552 blobs in 242 paths could be traced, leading to a large statistical pool from which dynamics and thermodynamical properties were derived. In Figs.~\ref{fig4} and~\ref{fig5} we show histograms of velocities and accelerations resulting from the individual tracing of the blobs. 

The total velocities of coronal rain blobs could be calculated from Doppler and projected velocities (since they constitute an orthogonal system in the velocity space), combining the spectral and imaging information provided by CRISP. Panels \textit{a}, \textit{b} and {\it c} of Fig.~\ref{fig4} show, respectively, normalized histograms of the total, Doppler and projected velocities. The determination of the Doppler velocities is explained in section~\ref{method}. As the condensations fall more than one measurement of the velocity is made (Doppler and projected velocities, as explained in section~\ref{method}), allowing estimates of the acceleration along the paths. The histogram for the accelerations is shown in Fig.~\ref{fig5}. We obtain a broad distribution of total velocities, from slow motions of a few km~s$^{-1}$ to high velocity downflows of more than 150~km~s$^{-1}$, and a mean around 70 km~s$^{-1}$. The resulting accelerations are in average small with respect to the solar surface gravity (0.274~km~s$^{-2}$), with a mean of 0.0835~km~s$^{-2}$. Tails indicating strong acceleration ($>0.5~$km~s$^{-2}$) and deceleration $<-0.5~$km~s$^{-2}$ are also found. It is important to note, however, that material in loops will be subject to the effective gravity only, i.e. the gravity component along the loop. The change of the average effective gravity along an ellipse with respect to its ellipticity can be calculated easily as $\langle g_{eff}\rangle=\frac{2}{\pi}\int_{0}^{\pi/2}g_{\sun}\cos\theta(s) ds$, where $\theta(s)$ is the angle between the vertical and the tangent to the path and $s$ is a variable parametrizing the path. It is found that for a ratio of loop height to half baseline between 0.5 and 2, $\langle g_{eff}\rangle$ varies roughly between 0.132~km~s$^{-2}$ and 0.21~km~s$^{-2}$, values that are significantly larger than the observed average value. Similar values for velocity and acceleration have been reported from limb observations with {\it TRACE} \citep{Schrijver_2001SoPh..198..325S}, with {\it SOHO}/EIT and in H$\alpha$ from Big Bear Solar Observatory \citep{DeGroof05}, and with {\it Hinode}/SOT \citep{Antolin_2010ApJ...716..154A, Antolin_Verwichte_2011ApJ...736..121A}.

\begin{figure}
\epsscale{1.}
\plotone{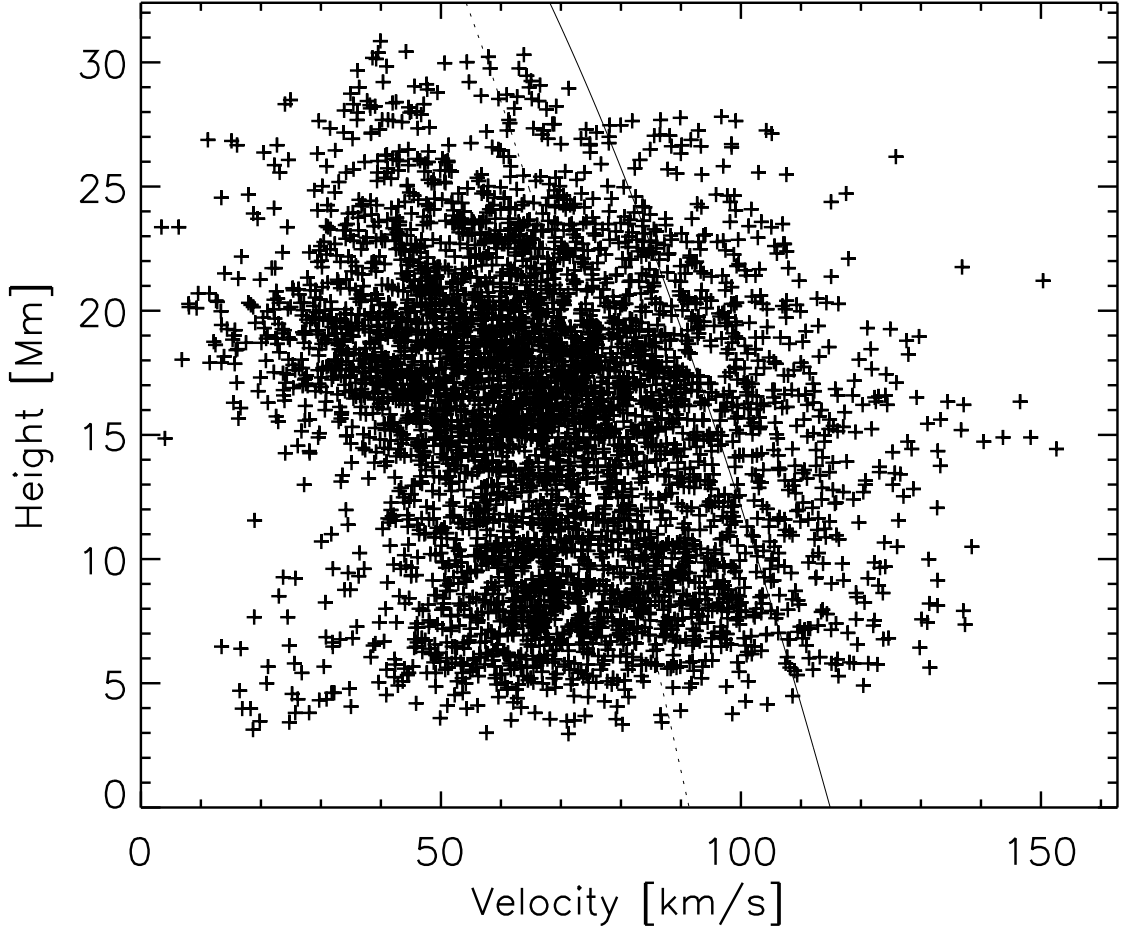}
\caption{Scatter plot of height versus the correspondent total velocity for off-limb measurements. The heights and velocities are calculated as explained in section~\ref{method}. For illustration purposes, the solid curve denotes the path that a condensation would follow if falling from a height of 50~Mm (an estimation of the height of a loop appearing in the \textit{Hinode}/EIS field of view, see section~\ref{rain}) and subject to an acceleration of 0.132~km~s$^{-2}$, the average effective gravity for a loop whose height to half baseline ratio is 0.5. The dashed curve denotes the same case but subject to the observed mean acceleration of 0.0835~km~s$^{-2}$. 
\label{fig6}}
\end{figure}

\begin{figure}
\epsscale{1.}
\plotone{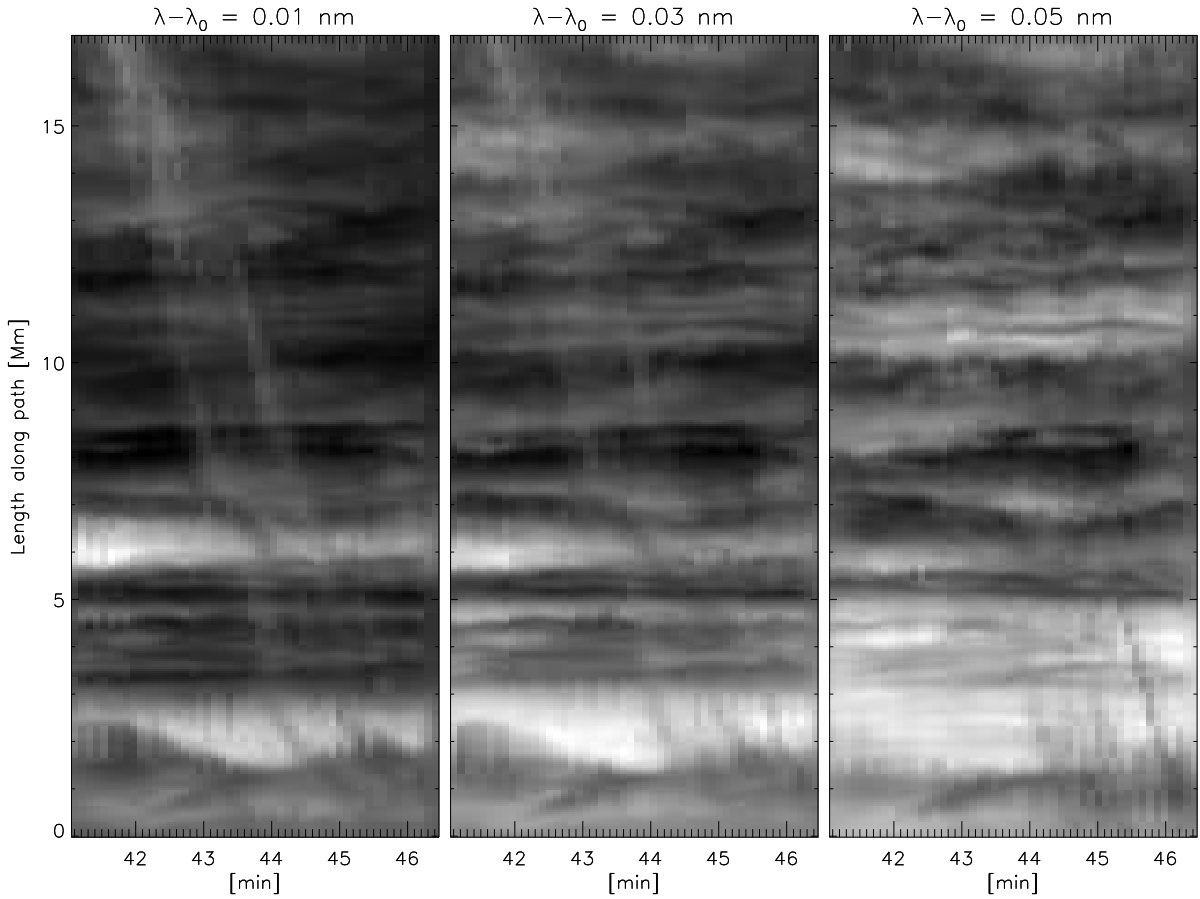}
\caption{Space-time diagram at different offsets from H$\alpha$ center (specified at the top of each panel) along a condensation path (off-limb case). Several condensations can be observed, appearing bright higher along the path towards line center, and dark at the bottom of the path at larger Doppler offset further into the red wing. 
\label{fig7}}
\end{figure}

Since the velocities are generally height dependent, in Fig.~\ref{fig6} we show a scatter plot of height versus total velocity for off-limb measurements. Heights were determined according to the method described in section~\ref{method}. For illustration purposes, the solid curve in the plot denotes the path that a condensation would follow if falling from a height of 50~Mm (an estimation of the height of a loop appearing in the \textit{Hinode}/EIS field of view, see section~\ref{rain}) and subject to an acceleration of 0.132~km~s$^{-2}$, the average effective gravity for a loop whose height to half baseline ratio is 0.5. The dashed curve denotes the same case but subject to the observed mean acceleration of 0.0835~km~s$^{-2}$. We notice that most of the measurements are located to the left of the solid curve, supporting the previously result, namely, that the accelerations are lower than the average effective gravity along a loop.

Strong decelerations of the condensations are sometimes observed, especially at lower atmospheric heights. An example of this is shown in Fig.~\ref{fig7}, which displays time slices of falling condensations along one path at 3 different line positions in the red wing of H$\alpha$. Towards line center the upper parts of the path are brighter, while the lower parts become more noticeable (dark in absorption against the bright background) towards the wing, accounting for the geometry of the path and the correspondent change in Doppler velocity (see section~\ref{trajec}). The increment in the projected speed is clear in the upper part, reaching values between 80 and 100~km~s$^{-1}$. Close to the footpoints, however, the 3 condensations display an abrupt change in speed, decreasing to values between 20 and 30~km~s$^{-1}$ for the first two condensations (between $t=44$~min and $t=45$~min for the first, and between $t=45$~min and $t=46$~min for the second), and $\simeq$70~km~s$^{-1}$ for the third one (mostly visible in the right-most panel, at $t\simeq46$~min) . The resulting decelerations are close to $-1~$km~s$^{-2}$ and $-0.5~$km~s$^{-2}$ respectively.

\subsection{Shapes}\label{shapes}

\begin{figure}
\epsscale{0.9}
\plotone{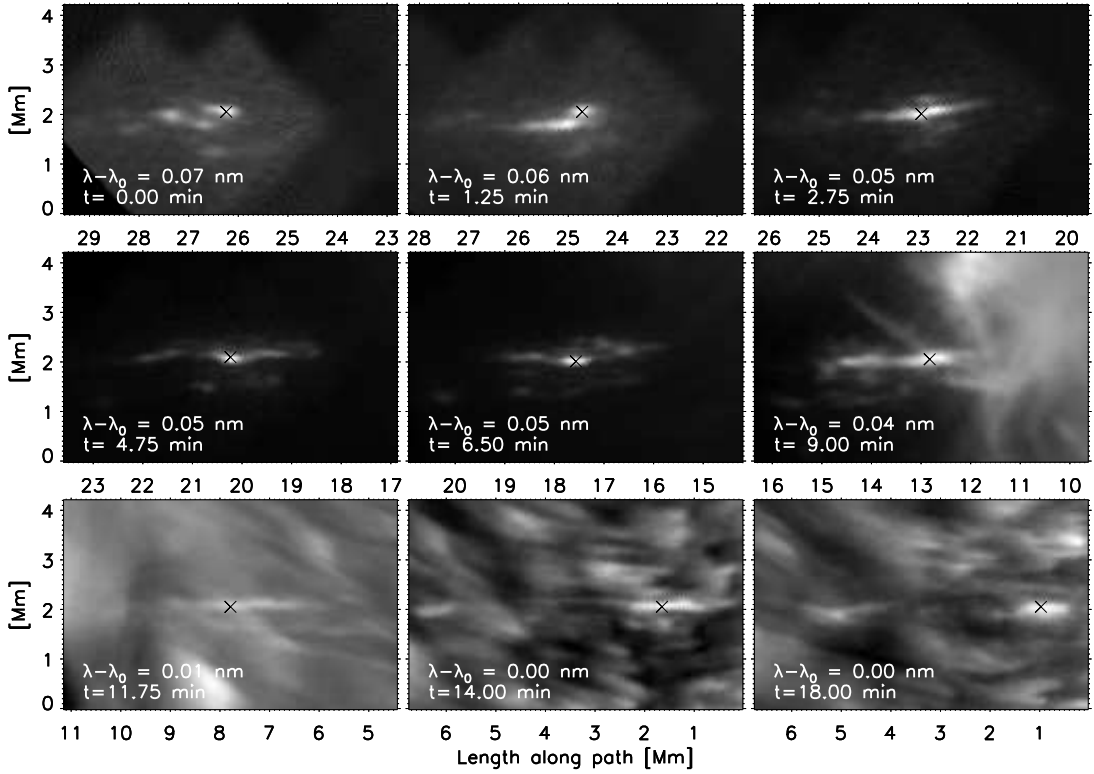}
\caption{Nine consecutive snapshots of a group of condensations, followed as they fall across the {\it SST} field of view towards the chromosphere. Each image corresponds to the brightest line position at which the blobs are observed. The offset from line center is specified in each panel, together with the time in minutes starting from the first image (top left). Notice the apparent continuous change in shape of the condensations, especially the elongation and separation. The images are obtained after application of the radial filter, as discussed in section~\ref{reduc}.
\label{fig8}}
\end{figure}

\begin{figure}
\epsscale{0.9}
\plotone{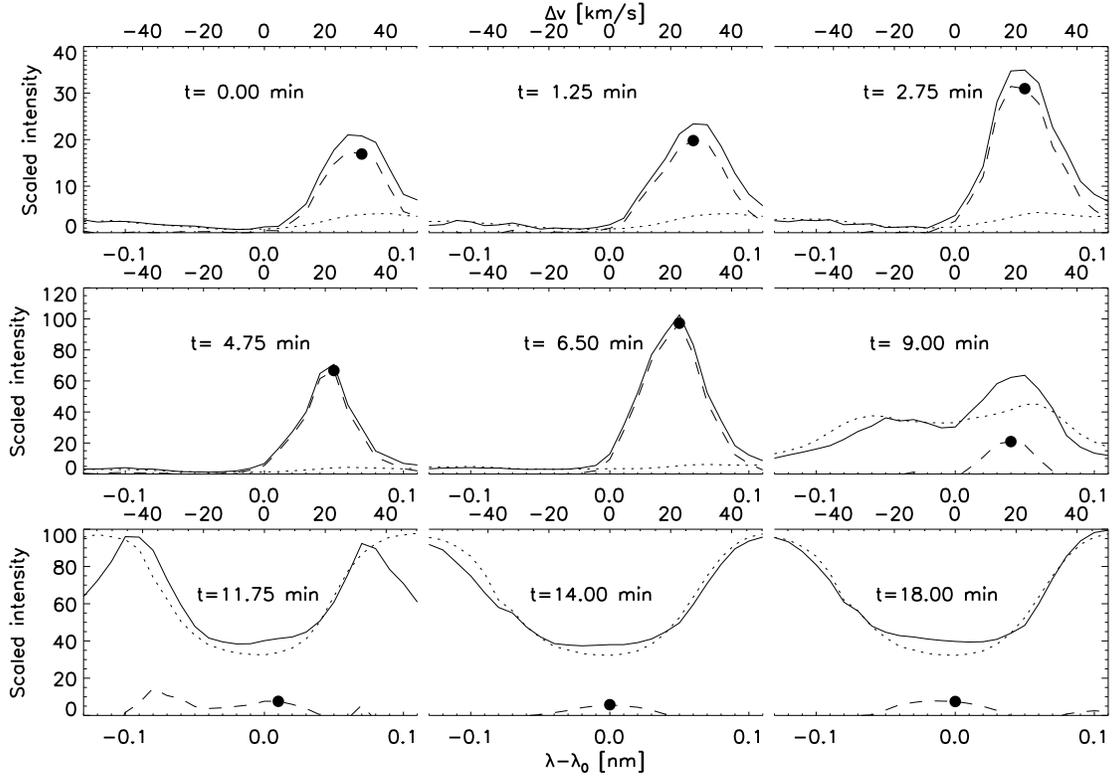}
\caption{The solid curve denotes  the line profile averaged over a small region around the maximum brightness of a blob, marked by the cross in Fig.~\ref{fig8}. The dotted curve denotes the average background line profile as calculated in section~\ref{method}. The dashed curve corresponds to the line profile of the condensation, defined as the difference of the two previous curves (such that the resulting profile is positive at the position of maximum blob intensity, specified by a black circle in each panel). The corresponding velocity for the offset from line center is specified in the upper axis of each panel. The line profiles are obtained after application of the radial filter to the images, as discussed in section~\ref{reduc}. The intensity scale corresponds to arbitrary units. 
\label{fig9}}
\end{figure}

One characteristic of the coronal rain observed here in H$\alpha$ appears to be the continuous change in shapes throughout the lifetimes of the condensations as they fall into the chromosphere. An example of this is shown in Fig.~\ref{fig8}, where we track a group of condensations from their first detection at a height close to 30~Mm, down to the chromosphere. Snapshots at 9 consecutive times (spanning 18~minutes) throughout the trajectory are displayed, selecting the wavelength position of maximum blob brightness at each time. A specific blob is chosen (black cross in the image), and in Fig.~\ref{fig9} we plot the shape of its line profile for each time averaged over a small region around the maximum brightness of the blob (solid curve). We also plot the average spectral profile of the background at each position along the path of the selected blob (dotted curve), as explained in section~\ref{method}. The dashed curve corresponds to the difference between both profiles (such that the resulting profile is positive at the position of maximum blob intensity). Note that the intensity scale in Figure~\ref{fig9} corresponds to arbitrary units, since the line profiles are obtained after application of the radial filter to the dataset, as discussed in section~\ref{reduc}. The most noticeable feature in the snapshots of Fig.~\ref{fig8} is the change in the overall shape of the condensations during the fall. In the first snapshot many blob-like condensations can be observed. However, in the second snapshot, only 1.25~minutes after, one elongated structure is mostly noticeable, together with our selected blob (with the black cross) and a few other fainter structures. In the next snapshot the elongated structure appears to have merged with our blob. In the following snapshots the elongation seems to increase until separation occurs (at $t=14~$minutes). The condensations then seem to decelerate considerably at a height of $\simeq0.5-2$~Mm. Although it is possible that seeing effects play a role in the visual determination of blobs, especially at these small scales, we believe these rapid morphological changes to be real due to their general occurrence throughout the observational sequence. The increase of length as the blobs fall seems to be common. On the other hand, we found in general little changes in the widths of the condensations with height. However, in some cases in which the condensations show considerable thickness in the corona ($\gtrsim0.7~$Mm), they become thinner towards the footpoints. 

\begin{figure}[!ht]
\begin{center}
$\begin{array}{c@{\hspace{-0.2in}}c@{\hspace{-0.2in}}c}
\includegraphics[scale=0.55]{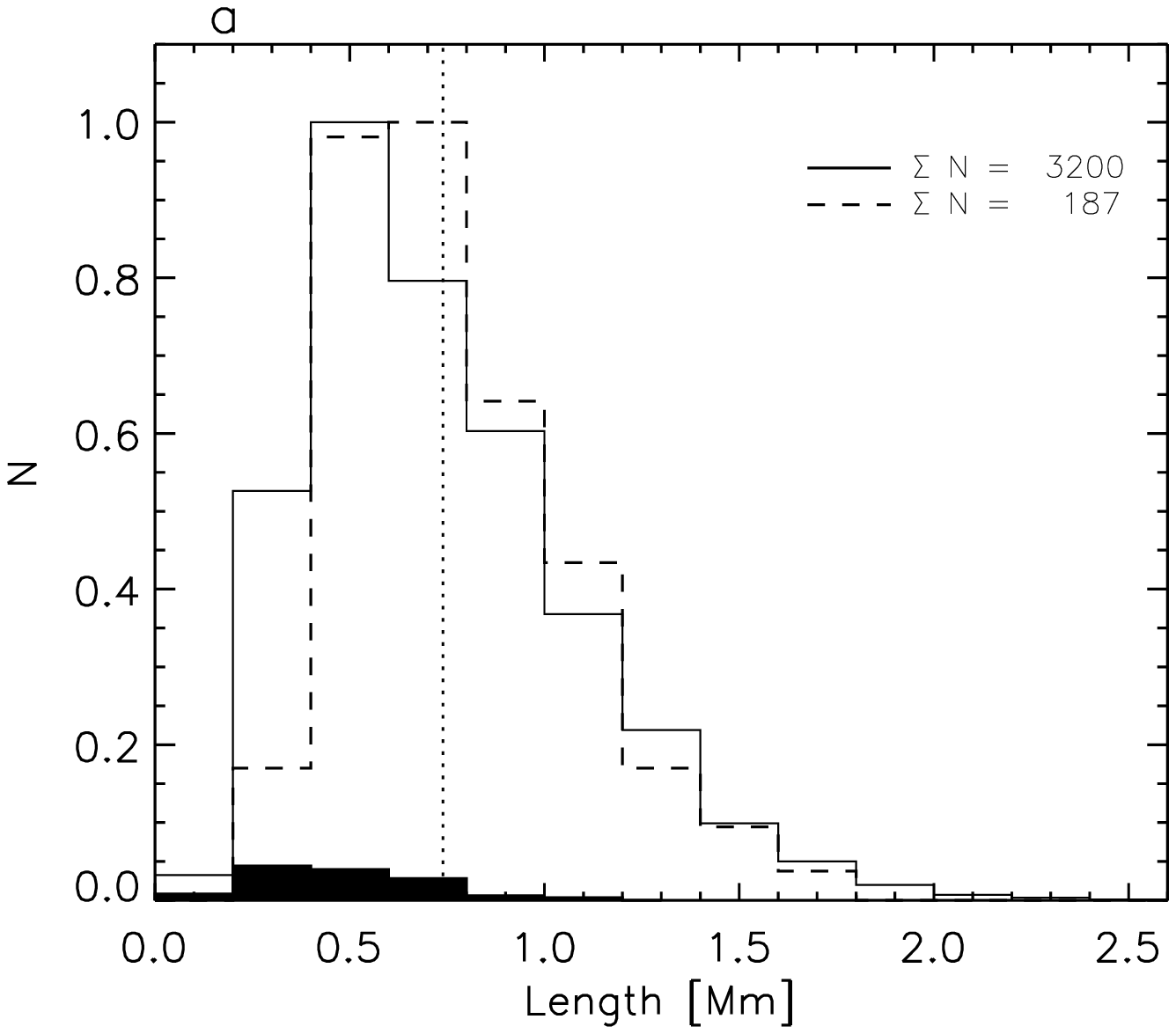}  &  
\includegraphics[scale=0.55]{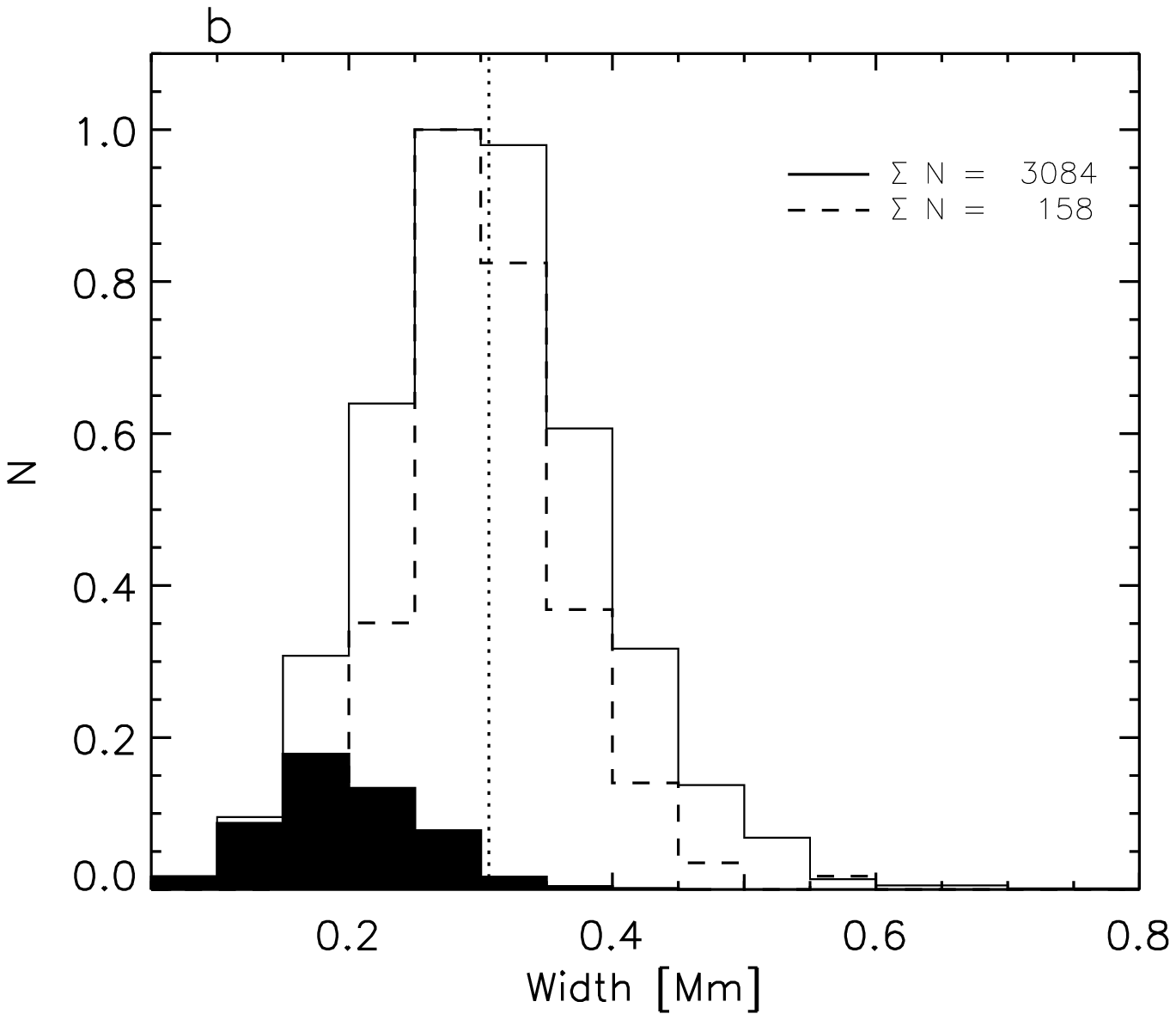}   \\
\end{array}$
\caption{Normalized histograms for the lengths (panel {\it a}) and widths (panel {\it b}) of the condensations. The solid and dashed lines correspond to off-limb and on-disc blobs respectively. The total number of measurements is specified in each panel. The dotted line corresponds to the average value over all measurements. The black histograms denote the measurements for which the 1-$\sigma$ errors in the gaussian fit are above $10~\%$ of the measured values (see section~\ref{shapes} for details).
\label{fig10}}
\end{center}
\end{figure}

\begin{figure}
\epsscale{.4}
\plotone{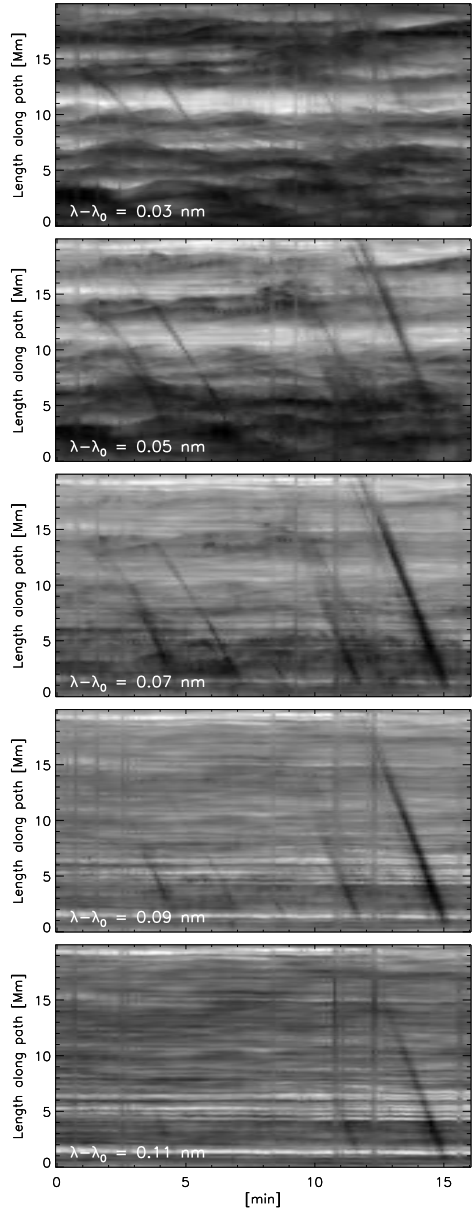}
\caption{Space-time diagram at different offsets from H$\alpha$ center (specified in the bottom left corner of each panel) along a path of an on-disc case. Several condensation traces are visible in absorption. By increasing the offset towards the red wing from line center the visible parts of the trajectories are shifted towards lower heights, accounting for the curvature of the path.
\label{fig11}}
\end{figure}

Histograms in Fig.~\ref{fig10} show the general picture for the lengths and widths of the observed condensations in H$\alpha$. Lengths present an average of $\sim1.5$~Mm, with a tail going from 0.4 to $\gtrsim3$~Mm. Widths present an average of $\sim0.5$~Mm, with a tail going from 0.2 to 0.8~Mm. The lengths and widths were calculated through gaussian fits over the intensity profile, as explained in section~\ref{method}. The black histograms in both panels of Fig.~\ref{fig10} denote the measurements for which the 1-$\sigma$ errors in the gaussian fit are above $10~\%$ of the measured values. 

\subsection{Trajectories}\label{trajec}

Having spectroscopic data additionally to imaging data allows an approximate 3D reconstruction of the trajectories of the blobs. For instance, in Fig.~\ref{fig9}, we note that the red shift in the line profiles of the condensation decreases throughout the trajectory from $\simeq32~$km~s$^{-1}$ to 0. The decrease in Doppler velocity of the blobs is more drastic close to the footpoints, where the chromospheric fibrilar structure appears in the background. On the other hand, the observed on-disc blobs close to the bottom right corner of Fig.~\ref{fig2} exhibit Doppler velocities that increase as the blobs get down to the chromosphere. This can be easily explained by a curvature effect of the paths, becoming more (less) perpendicular to the line of sight close to the footpoints for the off-limb blobs (on-disc blobs). An example of this effect for on-disc blobs is shown in Fig.~\ref{fig11}, where we show time slices along a path at different wavelength positions. See for instance the event whose trajectory terminates at $t=15$~min (right-most in the panels). As the wavelength is shifted further towards the red wing (from upper to bottom panel) different parts of the blob's trajectory become more noticeable, starting from the upper part and ending in the lower part towards the footpoints. The same tendency can be seen in the panels for various blobs along this path, suggesting an increase of the angle between the line of sight and the trajectory.

Using the MSDP imaging spectrograph at the Bia\l{l}k\'{o}v Observatory, 3D reconstructions of blob paths detaching from prominences have been done by \citet{Zapior_2010SoPh..267...95Z}. Since the set of projected and Doppler velocities $(\overrightarrow{v_{{\rm proj}}}, v_{{\rm Dop}})$ defines an orthogonal system, by determining the triad at each time, it is possible to reconstruct the path of each blob. In the case of coronal rain, assuming that H$\alpha$ blobs will, in general, follow the topology of the magnetic field, elliptic paths for each blob can be expected (see discussion in section~\ref{discuss1}). It is then possible, in theory, to retrieve the full geometrical shape of the loops, thus offering a direct way for mapping the coronal magnetic field. However, the errors involved in this reconstruction for the present data set are generally large, due mainly to the relatively short lengths over which the blobs can be traced with respect to the estimated full lengths of the loops. Furthermore, the paths of the H$\alpha$ blobs observed here do not correspond to perfectly straight or smoothly curving trajectories, effects mainly due to the seeing. Also, we cannot discard a possible lack of coupling at small length scales of the neutrals to surrounding ionized particles, a scenario worth investigating with numerical codes allowing for partial ionization (see section~\ref{discuss1}). In this section we thus do not attempt to reconstruct the full 3D paths but present estimates of the angles involved, which allows us to form a more complete picture of the surrounding corona.

\begin{figure}
\epsscale{.7}
\plotone{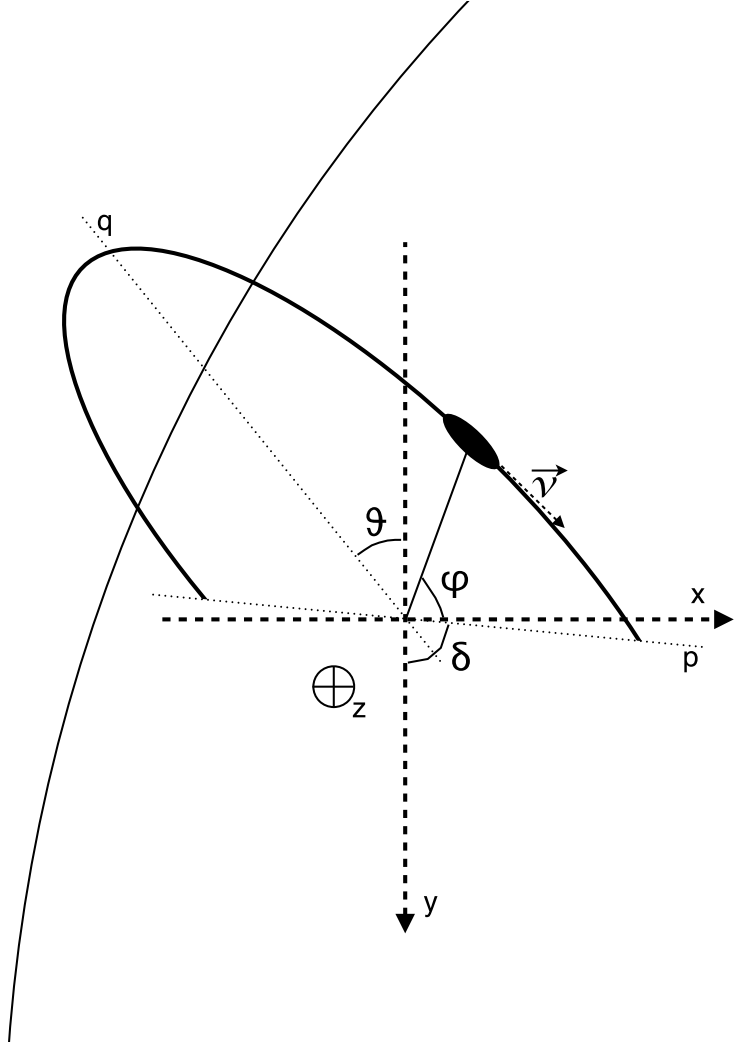}
\caption{A sketch showing the angles involved in the calculation of the trajectories of the blobs. The orientation and position of the loop is drawn according to the general geometry of the loops and bipolar regions observed with {\it Hinode}/EIS (Fig.~\ref{fig3})images and MDI magnetograms (see section~\ref{rain}). We take the $(x,y,z)$ coordinate system as indicated on the figure, where the origin is taken as the midpoint between the footpoints along the baseline of a given loop, and where the line of sight is along the $+z$ axis. $p$ and $q$ denote the baseline and semi-major axes of the loop. The angle $\delta$ corresponds to the angle between the projection of $p$ on the $x-y$ plane and the $y$ axis (counterclockwise with $\delta=0^{\circ}$ being the $+y$ direction). Similarly, $\theta$ corresponds to the angle between the projection of $q$ and the $y$ axis. $\varphi$ is the falling angle of the blob with respect to $p$ in the plane of the loop. We define also $\psi$ and $\gamma$ as the angles between $p$ and $-z$, and between $t$ and $- z$, respectively ($\psi=0$ or $\gamma=0$ denoting the $-z$ direction).
\label{fig12}}
\end{figure}

Determining a 3D loop trajectory involves the determination of various angles, as shown in Fig.~\ref{fig12}, where we have sketched a falling blob with velocity $\overrightarrow{v}$ from a loop towards the limb. There, $p$ and $q$ denote the baseline and semi-major axes of the loop. The geometry of the loop that is shown is set according to the general geometry of the loops observed with {\it Hinode}/EIS (Fig.~\ref{fig3}) and the bipolar regions observed in the MDI magnetograms (see section~\ref{rain}). Let us take the $(x,y,z)$ coordinate system as indicated on the figure, where the origin is taken as the midpoint between the footpoints along the baseline of a given loop, and where the line of sight is along the $+z$ axis. The angle $\delta$ corresponds to the angle between the projection of $p$ on the $x-y$ plane and the $y$ axis (counterclockwise with $\delta=0^{\circ}$ being the $+y$ direction). Similarly, $\theta$ corresponds to the angle between the projection of $q$ and the $y$ axis. $\varphi$ is the falling angle of the blob with respect to $p$ in the plane of the loop. Let us take also $\psi$ and $\gamma$ as the angles between $p$ and $-z$, and between $t$ and $- z$, respectively ($\psi=0$ or $\gamma=0$ denoting the $-z$ direction). Then, for a red-shifted blob the velocity vector can be written as:
\begin{equation}
\overrightarrow{v}=\left(\begin{array}{c}v_{{\rm proj,x}} \\v_{{\rm proj,y}} \\v_{{\rm Dop}}\end{array}\right)=|v|\left(\begin{array}{c}\cos\varphi\sin\gamma\sin\theta-\sin\varphi\sin\psi\sin\delta \\\cos\varphi\sin\gamma\cos\theta-\sin\varphi\sin\psi\cos\delta \\\sin\varphi\cos\psi+\cos\varphi\cos\gamma\end{array}\right).
\end{equation}
For a blue-shifted blob the last 2 terms in the $x$ and $y$ components are positive and the $z$ component is negative. 
Since $v^{2}=v_{{\rm proj}}^2+ v_{{\rm Dop}}^2$ the following relation must be satisfied:
\begin{equation}
\cos(\delta-\theta)=\frac{\cos\psi\cos\gamma}{\sin\psi\sin\gamma}.\label{angdif}
\end{equation}
Let $a\equiv\cos(\delta-\theta)$. Then we can write:
\begin{equation}\label{dop}
\frac{v_{{\rm Dop}}}{v}=\sin\varphi\cos\psi+\frac{a\cos\varphi\sin\psi}{\sqrt{\cos^{2}\psi+a^{2}\sin^{2}\psi}}.
\end{equation}
Since the observed loops are at the limb and the baselines are roughly directed along the East-West direction, we have $\psi\lesssim10^{\circ}$. Furthermore, we can assume the line of sight to be roughly in the plane of most loops (if the loops do not present significant inclinations with respect to the surface vertical). In this case, $\delta\simeq\theta$ and $a\simeq1$. Equations~(\ref{angdif}) and (\ref{dop}) then lead to:
\begin{eqnarray}\label{angfall}
\varphi+\psi&=&\sin^{-1}\left(\frac{v_{{\rm Dop}}}{v}\right)\\
\gamma&=&\frac{\pi}{2}-\psi.
\end{eqnarray}
In case of inclined loops and $\delta\neq\theta$ we expect $a\gtrsim\frac{1}{\sqrt{2}}$ since $0<\delta,\theta<\frac{\pi}{2}$ and therefore similar. To illustrate this, let $\psi=\epsilon_{1}$, $\gamma=\frac{\pi}{2}-\psi-\epsilon_{2}$. In the limit case of $a=\frac{1}{\sqrt{2}}$ allowing $\epsilon_{1}$ be up to $30^{\circ}$ we have $\epsilon_{2}<8^{\circ}$. In this case, assuming $\psi=10^{\circ}$ and $\frac{v_{{\rm Dop}}}{v}=0.3$ (a maximum value for most of our measurements) we obtain a difference of $\Delta\varphi\simeq3^{\circ}$ only.  

\begin{figure}[!ht]
\begin{center}
$\begin{array}{c@{\hspace{-0.2in}}c@{\hspace{-0.2in}}c}
\includegraphics[scale=0.55]{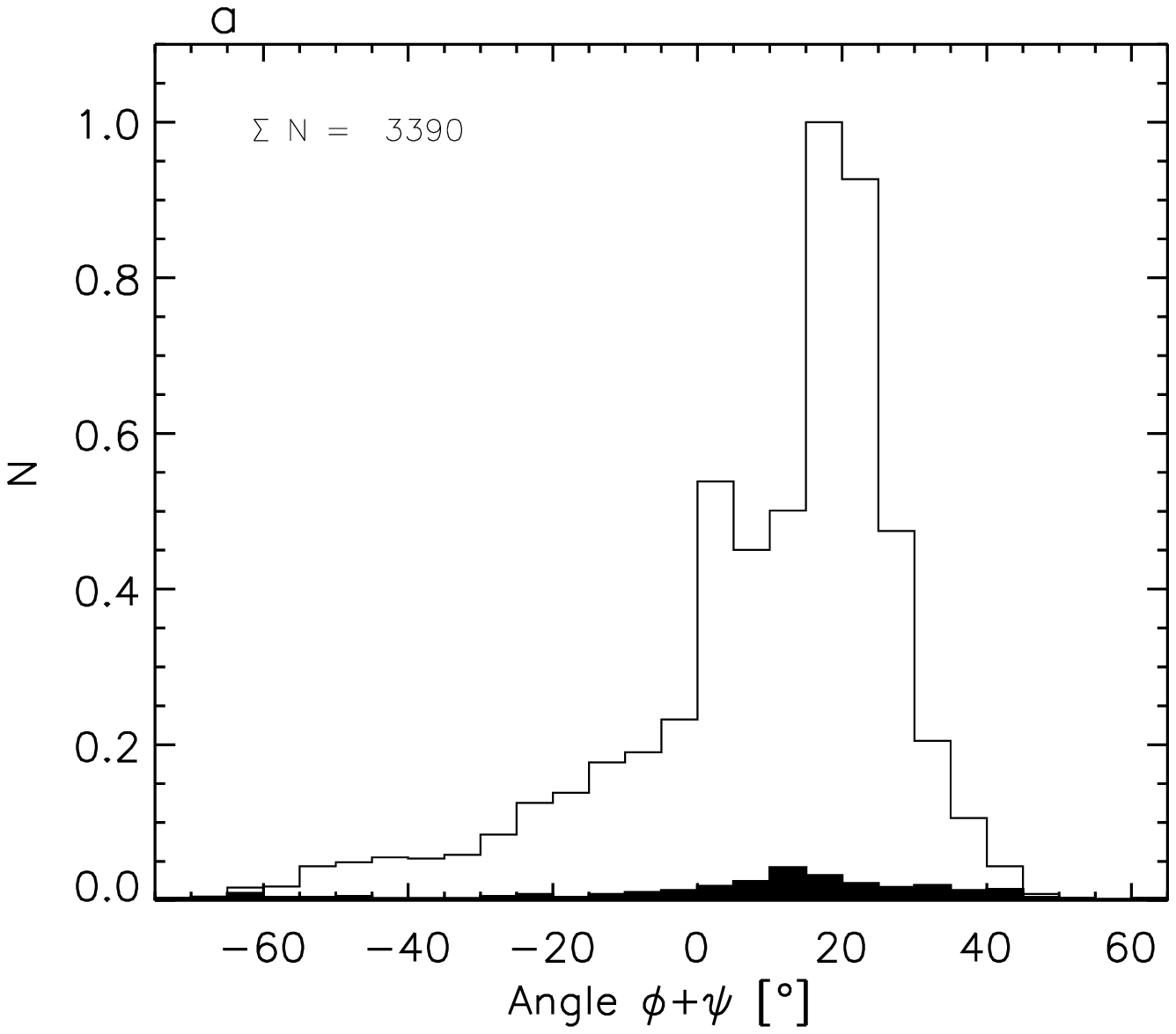}  &  
\includegraphics[scale=0.55]{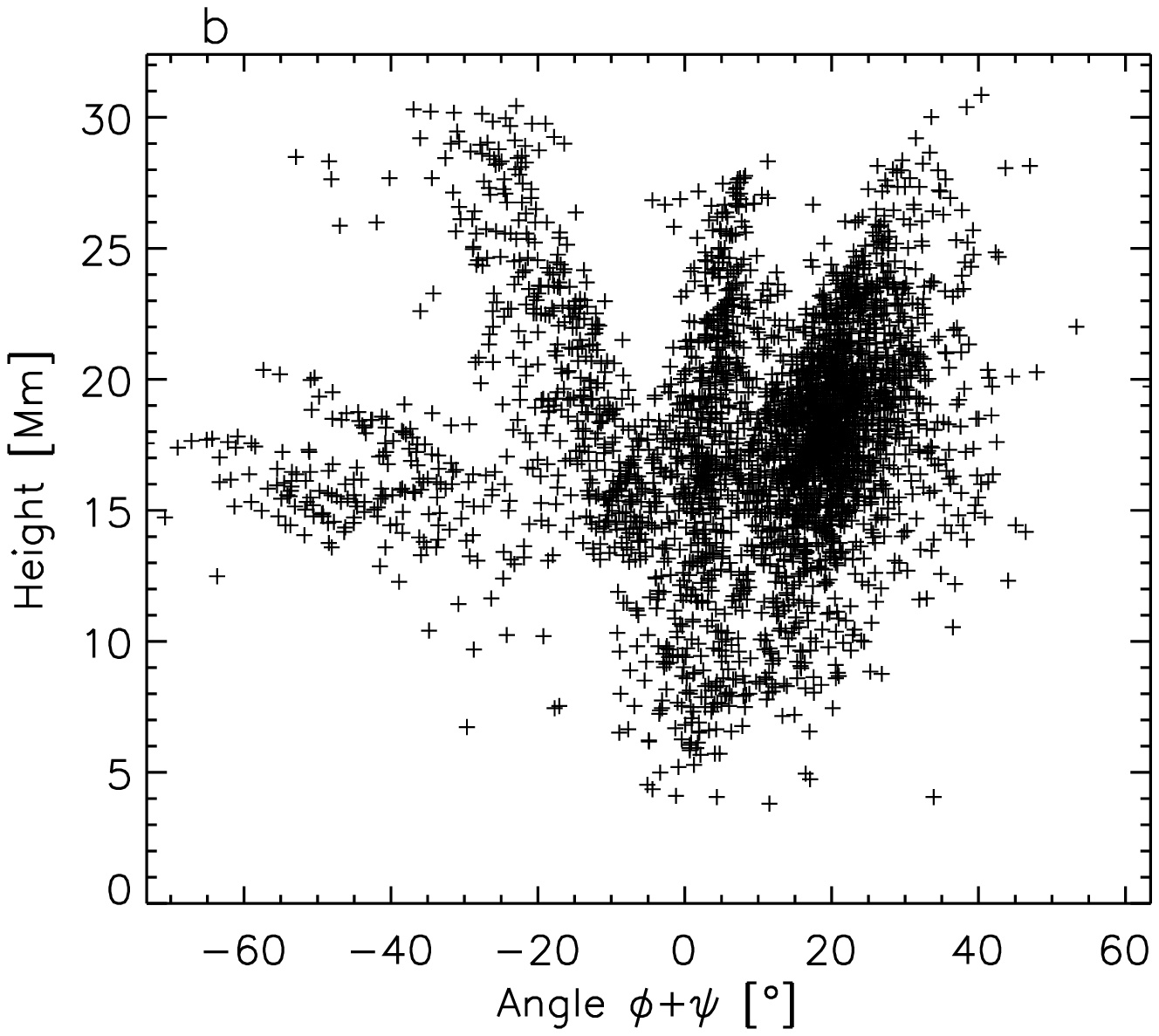}   \\
\end{array}$
\caption{In panel $a$ we show a normalized histogram of the $\phi+\psi$ angles (see Fig.~\ref{fig12}) calculated as explained in section~\ref{trajec}. The total number of measurements is specified in the top left corner. The black histogram denotes the angles for which the standard deviation is larger than $5^{\circ}$. In panel $b$ we show a scatter plot of the height of the blob versus the $\phi+\psi$ angle at that location. 
\label{fig13}}
\end{center}
\end{figure}

Panel $a$ of Fig.~\ref{fig13} shows the histogram corresponding to $\phi+\psi$ derived from Eq.~(\ref{angfall}). The negative and positive values in the plot correspond, respectively, to angles derived from blue-shifted and red-shifted blobs. The plot is composed of essentially 4 parts. Most of the blobs are red-shifted and show falling angles around $20^{\circ}$. A second less numerous group clusters around $\lesssim5^{\circ}$. Blue-shifted blobs show a small predominance around $-15^{\circ}$, with a long evenly distributed tail down to $-60^{\circ}$. In black we have denoted the angles for which the standard deviation is larger than $5^{\circ}$. 

\begin{figure}
\epsscale{1.}
\plotone{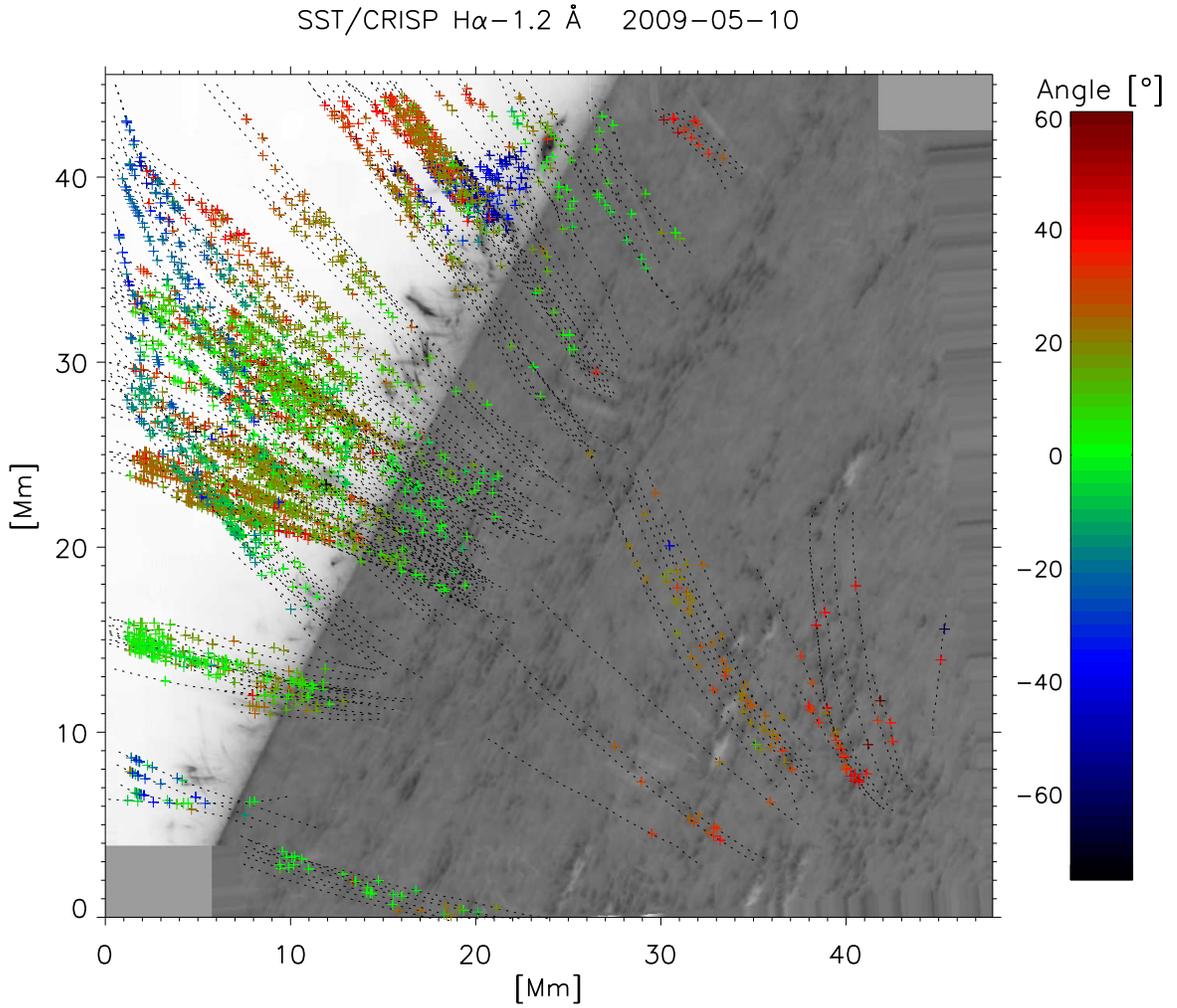}
\caption{Location of the $\phi+\psi$ angle measurements along each path (in dotted curves) over the full field of view (see section~\ref{trajec}). The black and white image corresponds to an offset from H$\alpha$ center towards the blue wing ($-1.2~$\AA), shown in inverted colors. 
\label{fig14}}
\end{figure}

Panel $b$ of Fig.~\ref{fig13} shows the heights of the blobs with respect to the angles. In order to reduce the errors due to projection in the calculation of heights we have excluded the on-disc events. Here also we can note 4 different groups, matching the groups in the histogram. We can see that in our field of view we have at least 4 big groups of loops with different geometries. The decrease of angle with height in both figures is clear, accounting for the loop-like paths of the blobs. In order to locate these groups in the {\it SST} field of view we have plotted in Fig.~\ref{fig14} over a full field of view the angle values at the corresponding height where the measurement is performed along each path. Notice the general decrease in angle with height and the location of the different loop families. The off-limb region in the middle left of the figure shows a superposition of 3 different loop families, 2 displaying redshits and 1 with blueshifts. 

\subsection{Temperatures}

A large unknown about coronal rain is the plasma state of the condensations of which it is composed. As generally agreed, this phenomenon is probably caused by thermal non-equilibrium (catastrophic cooling). Hence, its thermodynamic state should be determined partly by the characteristics of the radiative instability. The same mechanism is considered as a main agent in prominence formation, although the magnetic field topology in prominences appears to be very different (and thus may play a different role). It is thus possible that the thermodynamic state of these two phenomena is similar. This is supported by the fact that both phenomena have been observed in the same spectral lines. On the other hand, since prominences are long-lived structures, the neutral material may have more time to diffuse across the magnetic field, setting the possibility for different element population (and effects such as ambipolar diffusion). The thermal instability in the catastrophic cooling mechanism sets in when the plasma temperature is sufficiently low locally to trigger recombination of elements. A local loss of pressure accompanies this process, which triggers flows and creates the condensations. As the density rapidly increases so do the radiative losses, which lower the temperature even further. This cycle can be thus seen as an accelerating cascade towards lower temperatures whose timescale is set by the radiative timescales (in the order of minutes) and whose limits are set, apart from the limited lifetime of the blobs in the downfall, by possible heating mechanisms, such as radiative heating from the surrounding hot corona, wave heating, or other magnetic kinds of heating. 

\begin{figure}
\epsscale{1.}
\plotone{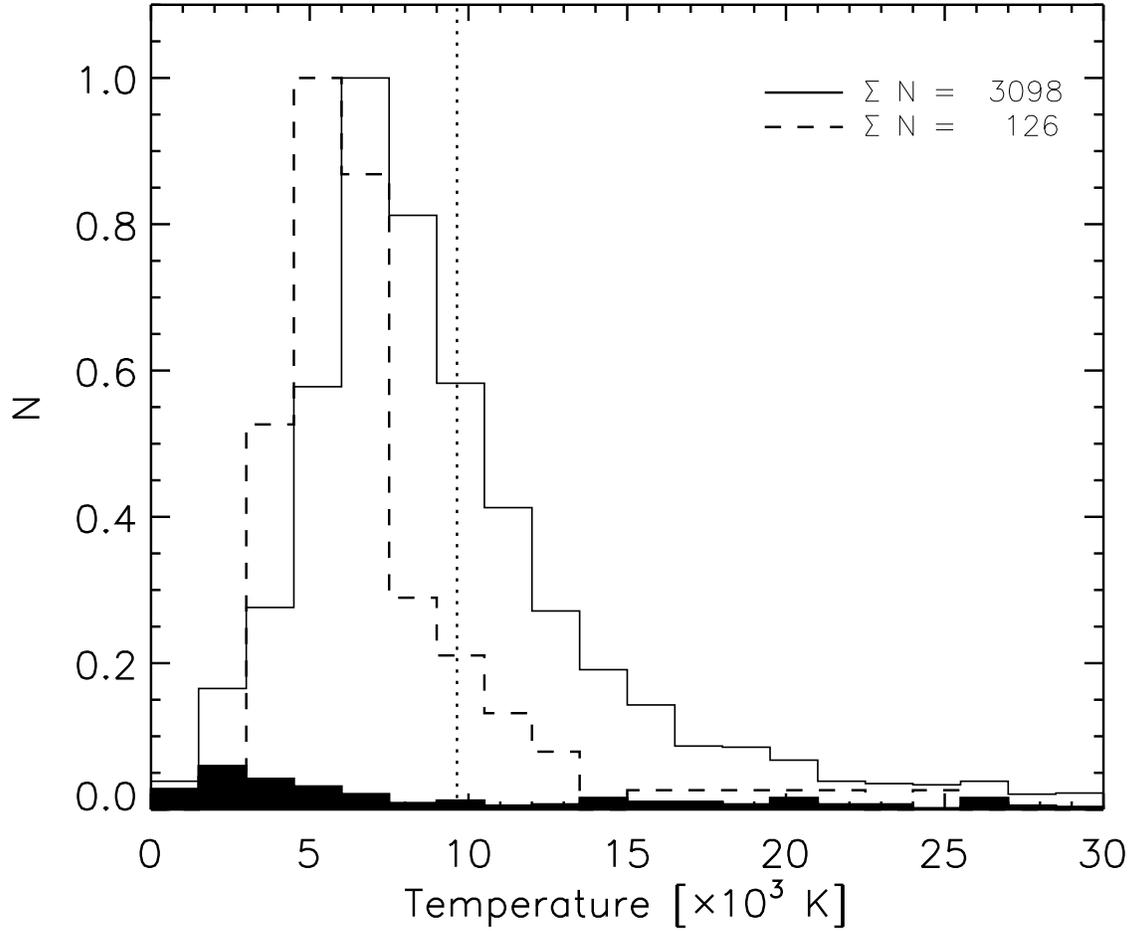}
\caption{Normalized histogram of the temperature measurements of the blobs according to Eq.~(\ref{tfwhm}). The solid and dashed lines correspond to off-limb and on-disc blobs respectively. The total number of measurements is specified in the top right corner. The dotted line corresponds to the average value over all measurements. The black histogram denotes the measurements for which the 1-$\sigma$ error estimate is greater than 10~$\%$ the calculated value. 
\label{fig15}}
\end{figure}

From the emission and absorption profiles of the condensations that we obtain through the procedure explained in section~\ref{method} we can estimate the plasma temperatures in coronal rain. The motion of coronal rain is mostly one-dimensional, directed roughly along the main loop axis, and unresolved internal plasma motions along the line of sight would be restricted to the small widths of the local strands. Hence, we can consider that most of the observed spectral line broadening comes from thermal motions. The FWHM obtained from gaussian fitting of the condensation's profile is then related to the temperature according to:
\begin{equation}\label{tfwhm}
{\rm FWHM}=2\sqrt{2\ln 2}\,\frac{\lambda_{{\rm 0}}}{c}\sqrt{\frac{2k_{{\rm B}}T}{m_{{\rm H}}}+v_{{\rm mic}}^2},
\end{equation}
where $\lambda_{{\rm 0}}$ denotes the H$\alpha$ line center, $m_{{\rm H}}$ is the hydrogen mass, $k_{{\rm B}}$ is Boltzmann's constant, $c$ is the speed of light and $v_{{\rm mic}}$ is the microscopic velocity accounting, for instance, for turbulence. Assuming that this last term is 0, we obtain upper bounds for the plasma temperature. In Figure~\ref{fig15} we show the histogram of the calculated temperatures for the off-limb (solid line) and on-disc (dashed line) condensations. The black histogram denotes measurements for which the 1-$\sigma$ error estimate is greater than 10~$\%$ of the calculated value. The mean temperature over all measurements is located around $10^{4}$~K. However, in both off-limb and on-disc cases we obtain a long tail towards higher temperatures extending up to $5\times10^{4}$~K, although for these cases the uncertainties increase as well. Rejecting this tail we can see that the mean is located around 7000~K and 5000~K for off-limb and on-disc cases respectively. Colder temperature plasma below 5000~K seems to exist as well.

\subsection{On-disc coronal rain}

As coronal rain falls from off-limb heights with a dark background down to the on-disc lower atmosphere it retains in most cases a bright emission profile, suggesting its dense and low temperature state characteristic of plasma which is chromospheric in nature. A natural question is then whether coronal rain can still be observed further into the disc. 
In the present data set, apart from the off-limb bright blobs, dark blobs further into the disc with similar characteristics as off-limb coronal rain are also observed. These cases correspond to the paths converging to the bright plage structures on the bottom right corner of Fig.~\ref{fig2}. Normalized histograms of the (total, Doppler and projected) velocities and accelerations for these on-disc cases are shown in the panels of Fig.~\ref{fig4} and in Fig.~\ref{fig5} in dashed curves. In general less than 10~\% of the total events correspond to on-disc cases, probably due to the small sizes and not large enough density and temperature contrast with respect to the background. Notice the similar shape in the distributions of these cases as for the off-limb coronal rain, which implies similar dynamics. Furthermore, these dark blobs also appear on short time scales along the loop and in a non-continuous way. In Fig.~\ref{fig10} and Fig.~\ref{fig15} we show in dashed lines, respectively, the lengths and widths, and the temperatures, for the on-disc blobs. We can see similar distributions as for the off-limb blobs. Since we have the same dynamics, morphology and thermodynamics we interpret these features as the on-disc counterpart of coronal rain. In Fig.~\ref{fig7}, as the spectral position is shifted from line center to the red wing of H$\alpha$, the condensations exhibit line profile shifts from emission to absorption, due to the increase in brightness of the background with the shift in line position. Our bright off-limb blobs thus naturally become dark depending on the background intensity. This example supports our interpretation of the dark blobs as coronal rain.

\section{Discussion}\label{discussion}

\subsection{Tracer of the internal structure of loops}\label{discuss1}

One of the most interesting features of coronal rain in our data is the very small sizes of the condensations it is composed of. While the lengths of these are in average $\sim1.5~$Mm and tend to increase as the blobs fall, their widths remain roughly constant during their fall, at $370~$km in average. The studied dataset presents good observational sequences with images close to the diffraction limit of the {\it SST} at H$\alpha$, allowing us to discern structure down to $\sim150~$km. For this reason, we expect that the most significant population of H$\alpha$ blobs is captured by the {\it SST}. As the blobs fall loop-like trajectories are traced, suggesting a tracing of the individual strands in loops, and thus supporting a multi-strand scenario. Following this line of thought, Fig.~\ref{fig2} does not show evidence for area cross section change of the loops with height, agreeing with previous observations of hot coronal loops in multiple filters of TRACE \citep{Klimchuk_etal_1992PASJ...44L.181K, Klimchuk_2000SoPh..193...53K, Watko_Klimchuk_2000SoPh..193...77W, Aschwanden_2005ApJ...633..499A, LopezFuentes_etal_2008ApJ...673..586L}. Little evidence of flux tube expansion at lower atmospheric heights was found. However, we expect that most of the magnetic field expansion should be located at photospheric heights, where the tracking of blobs is not possible. Furthermore, we find no evidence for strong braiding or twisting of the strands, whose scales are expected to be comparable to the 1~Mm scale of granulation. However, since the blobs do present significant changes of shape at smaller scales (on the order of 100~km) as shown in section~\ref{shapes}, and since we are not able to discern true transverse movement from seeing effects, it is possible that braiding and twisting exists at these small scales. Another possibility for explaining these small scale changes of shapes is a continuous change in the opacity of the material.

Let us estimate how good tracers of the magnetic field the observed H$\alpha$ blobs can be. We can identify two different processes that are important for the population of the upper level of the $n=3$ to $n=2$ transition that results in the H$\alpha$ spectral line. The first involves a radiative recombination process, in which a proton gains a free electron and experiences a de-excitation cascade down to the level $n=2$. This process depends on the availability of free electrons, slow enough to be caught by the protons and its rate is roughly $\nu_{rec}\simeq5\times10^{-13}n_{{\rm H^{+}}}~s^{-1}$ for a plasma at a temperature of 7000~K \citep{Seaton_1964PSS...12...55S}. It is a process which naturally offers tracing of the magnetic field. The second process requires the excitation of the neutral Hydrogen population, and if the coupling with protons is not strong then diffusion of particles across the magnetic field can take place. Collisions and charge exchange are the two processes that ensure a strong coupling between both populations, where charge exchange consists on one electron being exchanged between one proton and one neutral Hydrogen atom. According to \citet{Janev_1987ephh.book.....J} and \citet{Gilbert_2002ApJ...577..464G} the rate of such encounters between H and H$^{+}$ is roughly $\nu_{{\rm H^{+} H}}\simeq2\times10^{-8}n_{{\rm H^{+}}}~s^{-1}$ for a plasma with temperature of 7000~K. In this case, the mean free path for a neutral hydrogen atom results in $\lambda_{c}=\frac{v_{{\rm th}}}{\nu_{{\rm H^{+} H}}}\simeq6\times10^{13}/n_{{\rm H^{+}}}$ (in cgs units). In order to obtain a mean free path on the order of the blob size (a few 100~km) we would need a proton population of at most $n_{{\rm H^{+}}}\simeq10^{6}$~cm$^{-3}$. Considering chromospheric number densities of $10^{10}-10^{11}~$cm$^{-3}$ for the blobs, as usually found in prominences \citep{Hirayama_1985SoPh..100..415H} and also suggested from numerical simulations of thermal non-equilibrium \citep{Muller_2005AA...436.1067M, Antolin_2010ApJ...716..154A}, it implies an ionization fraction of $10^{-5}-10^{-4}$. The work of \citet{Heasley_Mihalas_1976ApJ...205..273H} shows that the radiative contribution from the photosphere, chromosphere and corona give ionization fractions that are higher by a factor of $10^{2}-10^{3}$. This suggests that coronal rain observed in H$\alpha$ corresponds to either protons recombining and/or to neutral Hydrogen strongly coupled to the ions, and in any case it is a good tracer of the coronal magnetic field, especially the internal structure of loops.

The relative motion of the condensations with respect to the solar surface can have a significant impact over the observed spectrum by altering the amount of incident radiation from the chromosphere and photosphere in the wavelength range of its line absorption profile \citep[an effect known as Doppler-brightening or Doppler-dimming,][]{Rompolt_1980HvaOB...4...39R,Rompolt_1980HvaOB...4...49R, Labrosse_etal_2007AA...463.1171L}. For instance, for prominence plasma at a height of 50~Mm with temperatures around $10^{4}~$K and electron densities between $10^{10}$ and $10^{11}$~cm$^{-3}$ the relative intensity in H$\alpha$ (defined as $E(v)/E(0)$, where $E(v)$ is the velocity dependent brightness) can be a factor of $2-3$ higher for velocities up to 200~km~s$^{-1}$. However, \citet{Heinzel_Rompolt_1987SoPh..110..171H} find that the effect on the electron densities is not important, and for velocities up to 150~km~s$^{-1}$ (as is the case of coronal rain) these are nearly constant. We should then not expect an important role of coronal rain kinematics (with respect to the incident radiation) on its ionization fraction. Main changes of the electron densities should then come from expansion and compression of the blobs only.

In theory, a direct ionization fraction measurement for a blob can be made if the integrated intensity over wavelength for the H$\alpha$ line is known. Indeed, \cite{Gouttebroze_etal_1993AAS...99..513G} and \cite{Heinzel_etal_1994AA...292..656H} show that a strong correlation exists, among others,  between the integrated intensity of H$\alpha$ (or the intensity at line center) and the emission measure, where the later is defined as ${\rm EM}=\int_{0}^{D}n_{e}^{2} dz$. Since we know the sizes of the emitting material ($D$) we can estimate the electron densities in the blobs, which would in turn give us an idea of the ionization fraction. However, the procedure for retrieving the absolute H$\alpha$ intensities is strongly affected by several factors from which an estimation of the error is difficult. The residual seeing effects, the instrumental stray light, the irradiation variation with solar altitude and the center-to-(off)limb variation of the absolute intensity are such factors that prevent a rigorous treatment.

When asserting that coronal rain is a good tracer of the coronal magnetic field we assume implicitly that the formation of coronal rain does not affect the general geometric structure of the loops, in the sense that the geometry of the strands suggested by the blobs will be the same as in the absence of them. Since the blobs are chromospheric in nature it is possible however that they have sufficient inertia to perturb the environment. In the case of prominences, for instance, it is still a matter of debate whether magnetic dips are present prior to their formation, or whether they are produced by the heavy chromospheric plasma. In an on-going work this situation is considered for the case of coronal rain, where the generation of transverse MHD waves by coronal rain is analyzed, as well as the influence of such waves on the kinematics of coronal rain \citep{Verwichte_etal_2011}. Preliminary results show that Alfv\'enic waves produced by a blob would generally have a short wavelength compared to the loop, but in case of fast and dense coronal rain (long loops), we could have a wavelength on the order of the loopÕs length (especially if the coronal magnetic field is not so strong). For short loops, 3D MHD numerical simulations by \citet{Zacharias_2011AA...532A.112Z} have shown that cool and dense blob-like ejections along loops due to sudden energy releases towards the footpoints can modify the structure of the coronal magnetic field.

\subsection{Probe of the local thermodynamic conditions in loops}

Despite the trajectories of the blobs observed in H$\alpha$ with the {\it SST} (Fig.~\ref{fig2}) matching roughly the geometries of the loops appearing in the {\it Hinode}/EIS field of view of Fig.~\ref{fig3}, it is not possible to see any trace of the cool plasma with the latter instrument. The cool and dense chromospheric state of coronal rain provides a sharp contrast with the surrounding transition region/coronal temperature environment, which should make the coronal rain material distinguishably dark in absorption. Similarly, coronal rain observed with {\it Hinode}/SOT \citep{Antolin_Verwichte_2011ApJ...736..121A} could not be detected with {\it Hinode}/XRT. This is partly due to the very small sizes of the blobs as seen in Fig.~\ref{fig10}, especially concerning their widths of 370~km in average. The EIS (and XRT) spatial resolution of $2^{\prime\prime}$ is insufficient to observe even the biggest blobs. On the other hand, the sporadic showers such as that in Fig.~\ref{fig1} (see section~\ref{rain}) have sufficient spatial extent that they in principle could be resolved by these instruments.

Probably the main factor restricting the detection of coronal rain in hot temperature lines is the amount of hot material along the line of sight between the observer and the location of the blobs, since in hot lines we can only spot coronal rain as matter in absorption against a bright background. Judging by the small sizes of the blobs compared to the radius of the loop structures observed in EIS ($\gtrsim7$~Mm), thermal non-equilibrium (leading to bright H$\alpha$ cores) seems to be restricted radially to only part of the loop. It is thus possible that the blobs are always surrounded by hot diffuse material belonging to the same magnetic field structure. The long tail towards hotter temperatures found in Fig.~\ref{fig15} may support this scenario. Limb observations of coronal rain with {\it TRACE} by \citet{Schrijver_2001SoPh..198..325S} show condensations bright in the 1216~\AA\ and 1600~\AA\ filters while the loop remains visible in the 171~\AA\ filter, supporting a multi-temperature scenario for the loop. Similarly, \citet{Kamio_etal_2011AA...532A..96K} observe the cool plasma in AIA-304~\AA\ immediately followed by emission in AIA-171~\AA. 3D numerical simulations with high spatial resolution in the corona are required to further test this scenario (codes with adaptive mesh refinement may be necessary for this task).

In the observations by \citet{Schrijver_2001SoPh..198..325S} a time lag of 200~s in average is observed between the 1600~\AA\ filter and the 1216~\AA\ filter, with the latter trailing behind, suggesting a progressive cooling of the condensations along their lifetimes. \citet{DeGroof05} have observed coronal rain simultaneously with {\it SoHO}/EIT 30.4~nm and with Big Bear H$\alpha$. They noticed that coronal rain condensations show up first in EIT~30.4 nm and later in H$\alpha$, with an increasing intensity of the later while falling down, supporting the cooling scenario. Furthermore, they found the blobs in H$\alpha$ to be smaller and more compact than in EIT 30.4~nm, suggesting cool core structures. In the present observations we confirm that coronal rain can have cool cores with average chromospheric temperatures of 7000~K. The true temperatures may be even lower if the Stark effect and turbulence are important. Some evidence for progressive cooling with decreasing height (or in time from the time of formation of the condensation) was found, but not substantial. This may be due to the fact that we are only observing the condensations in their latest stages of evolution due to the small field of view of the {\it SST}. 

1D numerical simulations of coronal rain suggest the presence of strong decelerations as the condensations fall to the chromosphere due to the compressed chromospheric plasma underneath. Furthermore, the transition region - corona layer is left oscillating up and down \citep{Muller_2004AA...424..289M, Antolin_2010ApJ...716..154A}. In our observations, strong deceleration below $-0.4~$km~s$^{-2}$ can only be observed in some cases such as that of Fig.~\ref{fig6} (see also the statistics of such cases in Fig.~\ref{fig5}). Furthermore, most of the observed decelerations only occur in the lower parts of the paths. Due to projection effects, it is difficult to locate the transition region - corona layer in our observations. If the background fibrilar structure observed partly in the off-limb region in H$\alpha$ towards line center is roughly on the same plane as the coronal rain, we can assume the transition region layer to be located towards the top of this structure. In this case, most of the observed decelerations (especially the strong ones) do not occur as soon as the blob enters the chromosphere, but happen in the lower brighter part of it (see Fig.~\ref{fig1}). The oscillations of the transition region plasma predicted by the numerical simulations in the works cited previously could not be observed at any height in any of our cases. 

The discrepancies between the present observations and the numerical results are partly due to the one-dimensional character of most of the models. In the simulations, the dense blobs occupy the entire width of the loop, and the compressed transition region plasma can only move in one dimension. It is then natural to expect strong decelerations and oscillations at higher atmospheric levels. In a 3D numerical simulation such oscillations may be absent or damped very quickly, or might occur only at lower atmospheric layers. Since the observed strong decelerations only occur in the lower chromospheric part, their detection during observations relies on being able to trace them to these heights, which constitutes a very challenging observation. Since we are not able to do so generally, it may explain the absence of the expected strong decelerations and subsequent bounce (oscillations). Another reason may be the highly dynamic character of the chromosphere (of which the fibrils are a good example), which would quickly erase any density perturbation caused by the falling blobs. 

Apart from gas pressure, magnetic pressure from transverse MHD waves has also been proposed as a mechanism to slow down coronal rain. In \citet{Antolin_Verwichte_2011ApJ...736..121A} observations with {\it Hinode}/SOT show such Alfv\'enic waves in a loop with coronal rain. Estimates of the upward wave pressure exerted by the wave seem to match the observed decelerations. In the present observations clear detection of transverse motion of the condensations is not possible due to seeing effects. Small shifts of the image are present from one line position to another which are not completely removed by the MOMFBD process, as explained in section~\ref{reduc}.

\subsection{Occurrence character of coronal rain - Do strands evolve independently?}

At first sight the on-line movie shows an ubiquitous character of coronal rain. In section~\ref{rain} we have also noted the sporadic appearance of large clumps of a few Mm in size, which we have denoted as `showers' (such as that shown in Fig.~\ref{fig1}). Given the sizes of these showers it is probable that the blobs observed in {\it TRACE} and EIT by \citet{Schrijver_2001SoPh..198..325S} and \citet{DeGroof05}, respectively, actually correspond to these showers. Now, a closer look at the movie reveals that, in general, the small blobs often appear in groups close together. Some of the groups are sufficiently close together and form showers. Let us analyze more closely the occurrence of coronal rain in the loops, especially, the occurrence of the phenomenon in neighboring strands.

\begin{figure}
\epsscale{1.}
\plotone{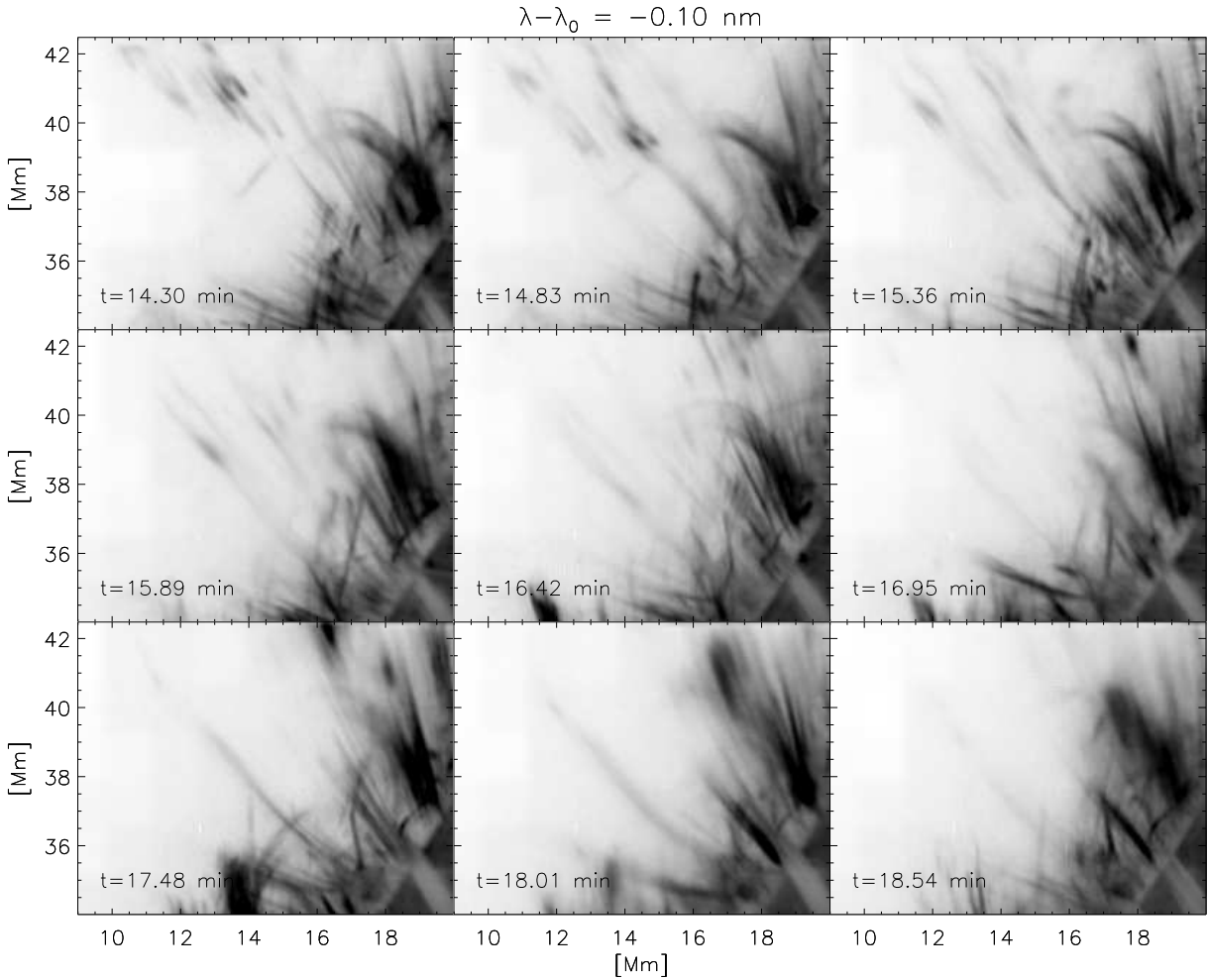}
\caption{Nine consecutive snapshots (in inverted colors) of falling condensations observed at an offset of $\lambda-\lambda_{0}=-0.1~$nm from line center. The times are specified in the bottom left corner of each snapshot. Notice that in each row we can see a group of condensations falling forming cool fronts across the loop. Notice also the dark-bright alternate pattern in the fronts, delineating strand-like structures in the loop.
\label{fig16}}
\end{figure}

Figure~\ref{fig16} shows 9 consecutive snapshots separated by 32~seconds from each other of a $\sim4\times5~$Mm region in our field of view (located in the top-middle part of Fig.~\ref{fig1}). A group of condensations can be observed to fall down diagonally from top left to bottom right, where the limb can be seen. The colors have been inverted for better visualization of the condensations. Notice that in each row we can see one group of condensations tracing the strands as they fall down and creating what appears to be `fronts' of condensations across the loop. Importantly, the condensations appear to be separated from one another, creating a bright-dark alternating pattern, thus highlighting the strand-like structure of the loop. The fronts are at least 4~Mm wide across and $1-2$~Mm in length. 

\begin{figure}
\epsscale{0.9}
\plotone{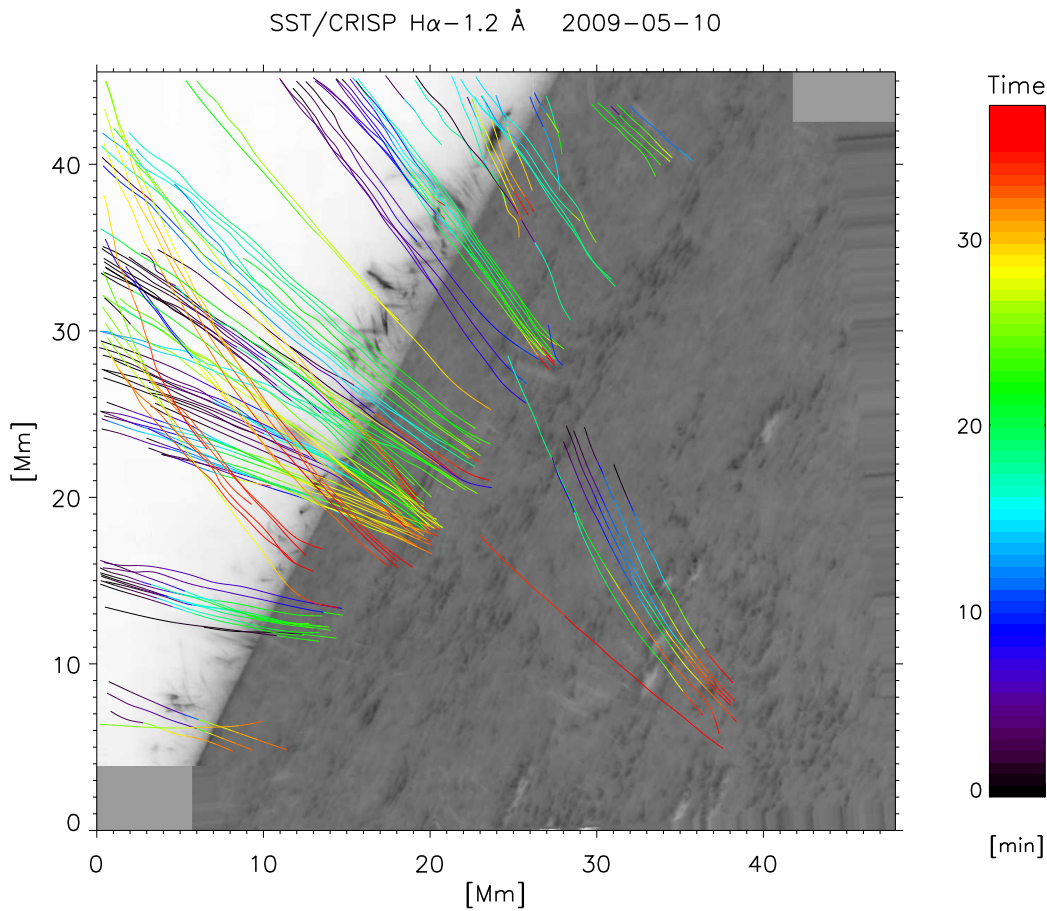}
\caption{Time occurrence of coronal rain for each strand over the full field of view. The color coding is done in the following way. Given a strand we first create a histogram of the time occurrence of the blobs in the strand (where a unit is given for each time step between the start and the end of the blob tracking). By then assigning a color to each time unit we color the strand so that the length of a specific color in the strand is proportional to its corresponding number in the histogram. For instance, the red strand in the bottom-middle of the figure has only blobs occurring at the end of the time sequence. For better visualization we only show the strands corresponding to the second data set, which spans 37~minutes and represents roughly two thirds of the total number of events. The black and white image corresponds to an offset from H$\alpha$ center towards the blue wing ($-1.2~$\AA), shown in inverted colors. 
\label{fig17}}
\end{figure}

The previous scenario appears to be rather common in the observed coronal rain. The general picture is presented in Fig.~\ref{fig17}, where we plot the strands over the field of view and where the color indicates the occurrence of the blobs in time. The color coding has been done in the following way. Given a strand we first create a histogram of the time occurrence of the blobs in the strand (where a unit is given for each time step between the start and the end of the blob tracking). By then assigning a color to each time unit we color the strand so that the length of a specific color in the strand is proportional to its corresponding number in the histogram. For instance, the red strand in the bottom-middle of the figure has only blobs occurring at the end of the time sequence. For better visualization we only show the strands corresponding to the second data set, which spans 37~minutes and represents roughly two thirds of the total number of events. Figure~\ref{fig17} shows clearly that neighboring strands tend to have the same color distribution, that is, same colors and same lengths for each color. In order to interpret this result correctly we have repeated the same procedure for a random time distribution of blobs for each path. Keeping the same number of blobs per path and the obtained minimum and maximum time span for a blob over the whole data we have generated uniformly-distributed pseudo random numbers for starting and ending times for each blob. The analog of Fig.~\ref{fig17} with such a procedure produces rainbow-like strands, where each color on the strand has in general the same length. On the other hand, Fig.~\ref{fig17} shows that most of the strands have only a few colors that are the same and have roughly the same length than those in the neighboring strands.

Figure~\ref{fig17} thus suggests that coronal rain occurs often simultaneously along neighboring strands. This implies an interesting scenario in which neighboring strands do not appear to cool independently, and probably meaning also that they are heated in similar ways. A puzzling consequence of this scenario is that throughout the lifetime of an observable loop (from heating to cooling, from being dense to being evacuated of plasma) strands may maintain a coherent state with similar temperature and density evolution. The coronal rain observed in \ion{Ca}{2}~H with {\it Hinode}/SOT reported in \citet{Antolin_2010ApJ...716..154A} and in \citet{Antolin_Verwichte_2011ApJ...736..121A} shows similar characteristics supporting these results. This does not mean, however, that such loops have a uniform temperature structure since each strand may be surrounded by diffuse hotter plasma, as the {\it TRACE} observations by \cite{Schrijver_2001SoPh..198..325S} and the {\it SDO}/AIA observations by \citet{Kamio_etal_2011AA...532A..96K} suggest. This scenario is also supported here since the coronal rain showers are not observed in {\it Hinode}/EIS. The thermal non-equilibrium mechanism may thus not act independently over single strands but on a significant number, probably a significant part of a loop, creating cool cores probably surrounded by diffuse hotter plasma. Co-observations between {\it SST} and {\it SDO}/AIA can further test this scenario.

The proposed scenario may pose a conflict with a general belief in the loop modeling community, namely, that all strands in loops evolve largely independently, and thus giving validity to one (or zero) dimensional models \citep{Klimchuk_2008ApJ...682.1351K}. Although transport coefficients are greatly reduced in the direction across the magnetic field, strands can still evolve coherently by sharing a common heating mechanism, as these observations suggest. A simple model explaining the observed behavior would be a heating mechanism concentrated towards the footpoints of the loop, imposing roughly the same heating scale height on neighboring strands.

\subsection{Is coronal rain a common active region phenomenon?}

One of the main results from numerical simulations regarding coronal rain has been to pinpoint the clear link between the heating scale height (among other parameters) and thermal non-equilibrium, and thus putting forward the now generally accepted mechanism behind coronal rain. Only loops with a relatively short heating scale height compared to the loop length seem to exhibit thermal non-equilibrium. The observational counterpart of this result is loops with coronal rain. In \citet{Antolin_2010ApJ...716..154A} it is shown that if torsional Alfv\'en waves have a predominant heating contribution through the process of non-linear mode conversion then thermal non-equilibrium is inhibited due to the resulting uniform heating from the waves. Together with the observational result of the previous section we can thus say with confidence that coronal rain has special relevance in the context of coronal heating. One of the most important questions that still needs a definitive answer is whether coronal rain is a common active region phenomenon or not. If it is then footpoint heating should be common, torsional Alfv\'en waves may not play an important heating role in active regions, and neighboring strands may have a similar thermodynamic evolution. 

Based on {\it TRACE} observations, \citet{Schrijver_2001SoPh..198..325S} estimated that less than 10\% of the pixels at a height of 10~Mm emit in the 1216~\AA\ filter at a given time, leading to timescales of $2-7$ days for a loop to exhibit coronal rain, and thus implying that coronal rain may not be common. However, cool downflows with coronal rain characteristics are often spotted in EUV active region studies \citep{Foukal_1976ApJ...210..575F, Foukal_1978ApJ...223.1046F, UgarteUrra_etal_2006ApJ...643.1245U, Oshea_etal_2007AA...475L..25O, Tripathi_etal_2009ApJ...694.1256T, Ugarte-Urra_etal_2009ApJ...695..642U, Landi_etal_2009ApJ...695..221L, Kamio_etal_2011AA...532A..96K}. As observed in the attached movie in this study, one of the main results of this paper is the relatively high abundance of coronal rain in the present active region, which however cannot be observed in {\it Hinode}/EIS nor XRT. Nor would {\it TRACE}, with a spatial resolution of $0.7-1$~Mm, be able to observe the biggest blobs. Such instruments would only be able to observe the sporadic shower events, which constitute the tip of the iceberg. A natural question is then whether coronal rain has been underestimated due to its small size. 

\begin{figure}
\epsscale{1.}
\plotone{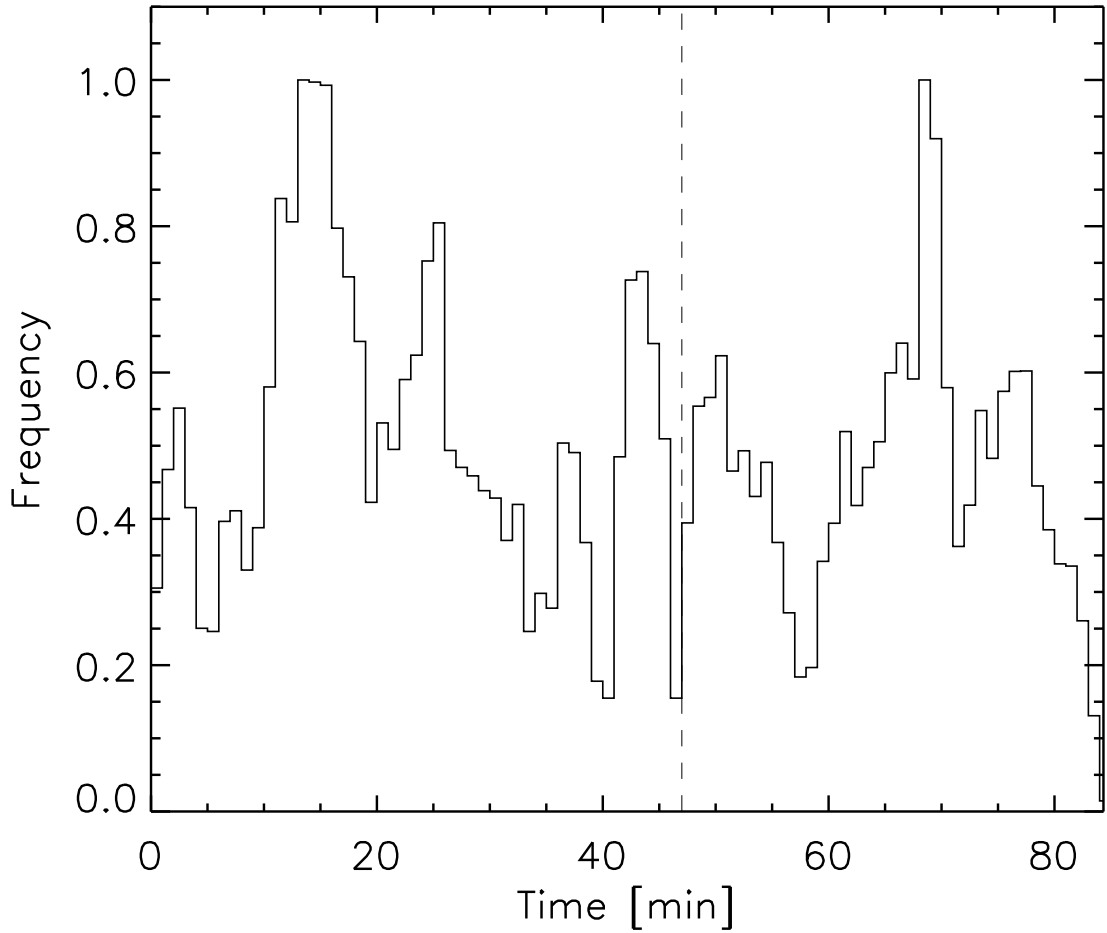}
\caption{Normalized histogram of the occurrence time of coronal rain over the total observational time of $\sim85~$minutes. Since the 2 data sets have a slightly different pointing, we have normalized each data set over their corresponding maxima, and present them next to each other in the figure in order to visualize the trends better). The dashed line in the figure denotes the separation between both data sets.
\label{fig18}}
\end{figure}

Let us estimate the fraction of coronal volume with coronal rain, and the corresponding mass downflow. Figure~\ref{fig18} shows a histogram of the occurrence time of the condensations over the total observational time of $\sim85~$minutes. Since the 2 data sets have a slightly different pointing (which may explain the different amount of coronal rain, the second data set having a larger off-limb region and exhibiting about $2-3$ times more coronal rain), we have normalized each data set over their corresponding maxima, and present them next to each other in the figure in order to visualize the trends better. The dashed line in the figure denotes the separation between both data sets. We can see that there are no clearly defined maxima, but many small peaks of coronal rain occurrence. On the other hand, numerical simulations indicate that the time span of a heating-cooling cycle produced by thermal non-equilibrium is very sensitive to parameters such as the heating scale height and the loop length. For loop lengths above 100~Mm and heating scale heights above 2~Mm the cycle times range from 1~hour to several days. Since coronal rain occurs in a roughly constant way in our observations, and the time span of these is relatively short compared to the estimated time range of the cycles it is reasonable to assume that each strand in our field of view undergoes a heating-cooling cycle at most once. Hence, we assume that any recurrent condensations in a strand separated by a significant amount of time (longer for instance than the travel time for a blob from loop apex to footpoint, i.e. roughly $15-25$~minutes for the present loops) occur actually in different (neighboring) strands but are seen in the same one due to projection effects. For the sake of simplicity, let us first assume that recurrent condensations in one strand occur in different neighboring strands, so that we assign only one condensation to each strand. Then the coronal volume with coronal rain is the volume occupied by strands that display coronal rain at least once during the full observation, and the number of such strands is just the number of observed condensations.

Notice that the main limiting factor for this calculation is the projection effect. Apart from the difficulty in determining whether condensations occur in the same or in neighboring strands, it is also difficult to estimate the depth along the line of sight to which the observed coronal rain corresponds to. Since we observe most of it falling straight (as shown by the angles in the panels of Fig.~\ref{fig13}), we approximate the depth as the size of the plage region below. Furthermore, let us assume that the blobs present an isotropic cross-sectional area in the perpendicular direction to their displacement. Then, the fraction of coronal volume with coronal rain can be approximated by:
\begin{equation}
{\rm Fraction}=f\frac{N_{s,cr}}{N_s},
\end{equation}
where $N_{s}$ is the total number of strands in the active region, $N_{s,cr}$ is the number of strands with coronal rain, and $f$ is the magnetic field filling factor. The total number of strands is given roughly by:
\begin{equation}
N_{s}=\frac{\mathcal{A}_{p}}{\mathcal{A}_{s}}\simeq\frac{\mathcal{A}_{{\rm proj},p}/\sin(\psi)}{\langle w\rangle^{2}},
\end{equation}
where $\langle w\rangle$ is the observed average width of the blobs, $\mathcal{A}_{s}$ is the cross-sectional area of a strand, $\mathcal{A}_{p}$ and $\mathcal{A}_{{\rm proj},p}$ are the true and projected areas of the plage region (on the plane of the sky) respectively, and $\psi$ is the angle between the line of sight and the tangent to the solar surface at the location of the plage. We estimate $\mathcal{A}_{{\rm proj},p}\simeq7~$Mm$\times22~$Mm$\simeq1.5\times10^{2}$~Mm$^{2}$. Since $\psi\simeq10^{\circ}$ (see section~\ref{trajec}) we have $\mathcal{A}_{p}=8.8\times10^{2}$~Mm$^{2}$. With a mean blob width average of 0.37~Mm we obtain $N_{s}\simeq6500$ strands. The total amount of blobs detected was roughly 2500. Assuming a filling factor of unity then the fraction of coronal volume displaying coronal rain is close to 40$\%$. On the other hand, if we assume that condensations in one strand recurring within 20~min correspond to the same strand then the amount of strands with coronal rain is reduced by a factor of 4, leading to a fraction of 10~$\%$. Note however that due to the manual procedure adopted in this work only blobs with enough contrast were chosen. We then expect the true fraction of coronal volume with rain to be between 10$\%$ and 40$\%$. In any case, given the relatively short time span of the current observations, the expected timescale for a loop in this active region to exhibit coronal rain is on the order of $4-15$ hours, a considerably lower estimate than that by \citet{Schrijver_2001SoPh..198..325S}. However, longer datasets with larger fields of view are needed to clearly answer the question of this section.

It is interesting to estimate the mass drain from the corona in the form of coronal rain. Let us assume that the blobs present a roughly isotropic shape in the plane perpendicular to their propagation. Their volume is then simply $V_{\rm blob}\simeq\langle w\rangle^{2}\langle l\rangle$, where $\langle w\rangle$ and $\langle l\rangle$ are, respectively, the average widths and lengths found in section~\ref{shapes}. Let us take a mass density for the blobs on the order of $10^{-13}$~g~cm$^{-3}$, corresponding to an average mass density often found in prominences \citep{Hirayama_1985SoPh..100..415H}, and also found in numerical simulations of coronal rain \citep{Muller_2005AA...436.1067M, Antolin_2010ApJ...716..154A}. Then, considering the total number of observed condensations and the total time of the dataset we obtain a mass drain on the order of $10^{10}$~g~s$^{-1}$. It is interesting to note that for the case of a prominence observed with {\it SDO}/AIA at 304\AA\  \cite{Liu_etal_11SPD....42.2119L} find a mass condensation rate of the same order, suggesting a general scenario in which thermal non-equilibrium does not depend on the coronal magnetic field configuration but rather on the heating of the magnetic field lines towards the footpoints. Based on average Doppler shifts in \ion{Ne}{8} and estimates of the areas for the observed blueshift paches \citet{Tian_etal_2009ApJ...704..883T} estimate the mass flux into the corona to be between $8\times10^{8}$ and $4\times10^{9}$ for a quiet Sun region. \citet{Pneuman_Kopp_1978SoPh...57...49P} and \citet{Beckers_1972ARAA..10...73B} estimate the particle flux into the corona from spicules to be $10^{15}$~cm$^{-2}$~s$^{-1}$. Taking the area of the plage for our active region as cited above this leads to a mass flux of $\sim1.5\times10^{10}$~g~s$^{-1}$. The amount of matter falling down as coronal rain is therefore substantial. This suggests a scenario in which a significant part of the spicular material injected into the corona \citep[which may be hot, according to][]{DePontieu_etal_2011Sci...331...55D} comes down as cool chromospheric material in the form of coronal rain, in agreement with the frequent cool downflows observed in EUV lines.

\section{Conclusions}\label{conclusions}

In this paper we have analyzed H$\alpha$ observations of coronal rain with the CRISP spectropolarimeter of the {\it SST}. The phenomenon is observed above an active region at the limb. Kinematics (total velocities and accelerations), shapes (widths and lengths), trajectories (angles of falling blobs) and thermodynamic properties (temperatures) were derived for the condensations. On-disc blobs dark in absorption were also analyzed and shown to have the same dynamical, morphological and thermodynamical properties as the off-limb blobs. We have further shown that the profile of off-limb blobs becomes an absorption profile when passing against a bright background,  from which we have concluded that on-disc blobs correspond to the same phenomenon.  

Being one of the first spectroscopic studies on coronal rain we have been able to measure the total velocities and shown that the falling velocities are lower than free fall, as has been suggested in previous works on the subject \citep{Schrijver_2001SoPh..198..325S, DeGroof_2004AA...415.1141D, Muller_2005ESASP.596E..37M}. The corresponding downward accelerations are lower than the effective gravity along elliptic loops, suggesting the presence of other forces than gravity, possibly gas pressure as suggested by most numerical simulations. Transverse magnetic waves could not be clearly detected due to seeing effects so we cannot discard their presence and influence on the dynamics of the blobs. The blobs could often be traced to chromospheric heights, close to what appears to be the footpoints of the loops that host them. Strong decelerations were observed in some cases, suggesting the increase of gas pressure at those heights expected from the higher local densities. 

Through orders of magnitude estimates we have shown that the coronal rain observed in H$\alpha$ is expected to come from a neutral Hydrogen population strongly coupled to the protons. At least on length scales on the order of the blob sizes (a few hundred km) no diffusion across the magnetic field is expected. Coronal rain can thus be considered as a tracer of the coronal magnetic field. Combining the projected and Doppler velocities we have shown that it is possible to retrieve the angles of fall of the blobs, allowing a reconstruction of the coronal magnetic field. No evidence for twisting or braiding of strands in loops was found, but we do not discard their existence at lower length scales (under 100 km), where turbulence may also be important. The tracing of strands by the blobs further suggest a constant area cross-section in the corona for these loops with no significant expansion down to chromospheric heights. The expected flux tube expansion may happen at lower (photospheric?) heights.

One of our main results concerns the sizes of the observed blobs. Lengths and widths on the order of 1.5~Mm and 0.37~Mm, respectively, were found, stressing the need of high spatial (and temporal) resolution to fully observe this phenomenon. The blobs further present average temperatures of 7000~K, and possibly even lower if turbulence and the Stark effect are important. Coronal rain thus present cool and compact chromospheric cores of a few 100~km in width. Some evidence of progressive cooling over time was found, as suggested from previous work. We further found that coronal rain occurs frequently in neighboring strands in a simultaneous way, forming groups of condensations which, if close enough together, are seen as large clumps. We have termed these sporadic events as `showers', and they can have widths up to a few Mm. This is probably what has been observed in the past with instruments of lower spatial resolution such as {\it TRACE} and {\it SOHO}/EIT. Since we did not observe here any evidence of cool blobs nor showers in the co-temporal {\it Hinode}/EIS observation we suspect that loops with coronal rain are multi-thermal, as suggested previously by \cite{Schrijver_2001SoPh..198..325S}. A scenario in which the cool coronal rain cores are surrounded by diffuse hotter plasma needs to be further investigated. 

Our observations support the multi-stranded loop scenario and suggest that a significant fraction of strands in a loop do not have an independent thermodynamic evolution. Indeed, neighboring strands often display a coherent cooling (in the form of coronal rain) of a significant number of strands in a loop, suggesting a heating mechanism which acts in roughly the same way on neighboring strands. A footpoint heating mechanism imparting a similar heating scale height over all or part of the strands in a loop is a simple scenario that can explain these observations. Since we expect an expansion of the loops at low (possibly photospheric) heights spatially small heating events may suffice for such purpose. 

A coherent evolution of strands in a loop throughout their lifetime (from heating to cooling, from being dense to being evacuated) is a puzzling scenario that needs further observational testing. In case of being a common scenario it  may pose several constraints on coronal heating models, especially on one-dimensional models. For instance, one constraint may be that a loop cannot be modeled by a collection of independently evolving one-dimensional strands. This may explain the disagreement regarding the importance of thermal non-equilibrium in the corona between the 1D \citep{Klimchuk_2010ApJ...714.1239K} and the 3D simulations \citep{Klimchuk_2010ApJ...714.1239K, Lionello_etal_2010AGUFMSH31C1811L}.

The present observations suggest that coronal rain may not be a sporadic phenomenon of active regions as previously thought, but a rather common phenomenon deeply linked to the heating mechanisms of coronal loops. We have estimated the fraction of coronal volume with coronal rain to be between 10~\% and 40~\%. This is however strongly dependent on the magnetic field filling factor and on the projection effects. We have further estimated the occurrence time of this phenomenon in a loop to be between 4 and 15~hours. Longer datasets with larger fields of view are needed to clearly answer the question on how common coronal rain is. The obtained mass drain rate in the form of coronal rain is roughly $10^{10}$~g~s$^{-1}$ taking a mass density of $10^{-13}$~g~cm$^{-3}$ for the condensations. This number is on the same order of magnitude as the estimated mass flux into the corona from spicules, reinforcing the idea that coronal rain is an important phenomenon. This suggests a scenario in which the hot spicular material injected into the corona falls back cool, `raining' down. 

Irrespectively of the occurrence time of coronal rain, we have shown in this paper the large potential that this phenomenon beholds in the understanding of the coronal magnetic field. In \citet{Antolin_2010ApJ...716..154A} we have shown that coronal rain can act as a marker of the coronal heating mechanisms. Here, besides further elucidating its link to coronal heating, we have shown that it constitutes a tracer of the internal structure of loops and that it can be a probe of the local thermodynamic conditions in loops.

\acknowledgments

The authors would like to acknowledge G. Vissers for the development of the versatile analysis tool CRISPEX, from which the analysis in this paper greatly benefited. We would like to thank M. Carlsson, V. Hansteen, P. Judge, E. Khomenko, M. Collados and J.C. Vial for the fruitful comments and discussions that lead to a significant improvement of the manuscript. We would further like to thank D. Sekse and Y. Lin who were co-observers at the {\it SST} on the day of the presented data. The Swedish 1-m Solar Telescope is operated on the island of La Palma by the Institute for Solar Physics of the Royal Swedish Academy of Sciences in the Spanish Observatorio del Roque de los Muchachos of the Instituto de Astrof\'isica de Canarias. 

\bibliographystyle{aa}
\bibliography{ms.bbl}  

\clearpage

\end{document}